\documentclass[aps,pra,superscriptaddress,nofootinbib,notitlepage,longbibliography]{revtex4-1}

\usepackage[english]{babel}
\usepackage[T1]{fontenc} 
 \usepackage{times}

\usepackage{amsthm} 
\usepackage{amssymb} 
\usepackage{amsmath} 
\usepackage{mathbbol} 
\usepackage[normalem]{ulem}
\usepackage[usenames,dvipsnames]{color}
\usepackage{graphicx}
\usepackage{overpic}
\usepackage{xifthen}
\usepackage{verbatim}
\usepackage{booktabs}
\usepackage{float}

\usepackage{paralist} 

\usepackage{tikz}

\usepackage{mathtools} 

\usepackage{hyperref}

\usepackage{listings}
\newcommand{\pyline}[1]%
{
\definecolor{number}{gray}{0.6}
\definecolor{keywords}{rgb}{1.0,0.3,0.3}
\definecolor{comments}{rgb}{0.1,0.65,0.1}
\definecolor{strings}{rgb}{0.3,0.0,1.0}
\lstset{language=python,
        morekeywords={switch,case},
        sensitive=true,
        showspaces=false, 
        basicstyle=\ttfamily\small\mdseries,
      	keywordstyle=\bfseries\color{keywords},
 	      commentstyle=\color{comments},
 	      stringstyle=\color{strings},
 	      numbers=left,
 	      numberstyle=\scriptsize\color{number},
 	      stepnumber=1,
 	      breaklines=true,
 	      frame=none,
        showstringspaces=false,
 	      tabsize=1,
 	      xleftmargin=0pt,
 	      xrightmargin=0pt,
 	      aboveskip=\bigskipamount,
      	belowskip=0pt
}
\lstinline{#1}
}

\newcommand{\pycode}[3]%
{
\definecolor{number}{gray}{0.6}
\definecolor{keywords}{rgb}{1.0,0.3,0.3}
\definecolor{comments}{rgb}{0.1,0.65,0.1}
\definecolor{strings}{rgb}{0.3,0.0,1.0}
\lstset{language=python,
        morekeywords={switch,case},
        sensitive=true,
        showspaces=false, 
        basicstyle=\ttfamily\small\mdseries,
      	keywordstyle=\bfseries\color{keywords},
 	      commentstyle=\color{comments},
 	      stringstyle=\color{strings},
 	      numbers=left,
 	      numberstyle=\scriptsize\color{number},
 	      stepnumber=1,
 	      breaklines=true,
 	      frame=none,
        showstringspaces=false,
 	      tabsize=4,
 	      xleftmargin=\bigskipamount,
 	      xrightmargin=\bigskipamount,
 	      aboveskip=\bigskipamount,
      	belowskip=0pt
}
\lstinputlisting[caption=#2, label=#3, frame=tb]{#1}
}
\newcommand{\pyfrag}[5]%
{
\definecolor{number}{gray}{0.6}
\definecolor{keywords}{rgb}{1.0,0.3,0.3}
\definecolor{comments}{rgb}{0.1,0.65,0.1}
\definecolor{strings}{rgb}{0.3,0.0,1.0}
\lstset{language=python,
        morekeywords={switch,case},
        sensitive=true,
        showspaces=false, 
        basicstyle=\ttfamily\small\mdseries,
      	keywordstyle=\bfseries\color{keywords},
 	      commentstyle=\color{comments},
 	      stringstyle=\color{strings},
 	      numbers=left,
 	      numberstyle=\scriptsize\color{number},
 	      stepnumber=1,
 	      breaklines=true,
 	      frame=none,
        showstringspaces=false,
 	      tabsize=4,
 	      xleftmargin=\bigskipamount,
 	      xrightmargin=\bigskipamount,
 	      aboveskip=\bigskipamount,
      	belowskip=\bigskipamount
}
\lstinputlisting[caption=#2, label=#3, frame=tb, firstnumber=#4, firstline=#4,
  lastline=#5]{#1}
}

\newcommand{\mioseventeenx}{\emph{2x(Intel Xeon E5-2680 Dodeca-core) 24 Cores 2.50GHz}} 

\definecolor{dg}{HTML}{006400} 

\newcommand{\Heff}{H_{\mathrm{eff}}}
\newcommand{\rqt}{r_{\mathcal{N}}}

\newcommand{\OSMPS}{OSMPS}

\newcommand{\miosixxeightx}{\emph{Penguin Relion 1752DDR/QSFP 2x(Intel X5675) 12 cores 3.06 GHz}}

\newcommand{\bpm}{\begin{pmatrix}}
\newcommand{\epm}{\end{pmatrix}}

\newcommand{\dt}{\mathrm{d}t}
\newcommand{\Tens}[4][]{\ifthenelse{\equal{#1}{}}
   {  
      #2_{#3}^{#4}
   }
   {  
      #2_{#3}^{\lbrack #1 \rbrack \, #4}
   }
}

\makeatletter
\newcommand{\pushright}[1]{\ifmeasuring@#1\else\omit\hfill$\displaystyle#1$\fi\ignorespaces}
\makeatother

\renewcommand{\i}{\mathrm{i}}
\newcommand{\1}{\mathbb{I}}

\newcommand{\ket}[1]{\left|{#1}\right\rangle}
\newcommand{\Lket}[1]{\left|{#1}\right\rangle\!\rangle}
\newcommand{\Lbra}[1]{\left\langle\!\langle{#1}\right|}

\newcommand{\bra}[1]{\left\langle{#1}\right|}

\renewcommand{\vec}[1]{\mathbf{#1}}

\definecolor{daniel}{rgb}{0,.1,1}

\definecolor{kenji}{RGB}{0,100,00} 

\definecolor{lincoln}{RGB}{40,180,40}

\definecolor{ulmcolor}{rgb}{.4,0,.6}

\definecolor{ulmmethod}{rgb}{0,.3,.6}

\definecolor{dangercolor}{rgb}{0.8,0.,0.}

\newcommand{\Tcpu}{T_{\mathrm{CPU}}}
\newcommand{\ezz}{\epsilon_{zz}}

\usepackage[normalem]{ulem} 
\usepackage{cancel}

\newcommand{\csm}{Department of Physics, Colorado School of Mines, Golden,
  Colorado 80401, USA}
\newcommand{\ulm}{Institute for Complex Quantum systems and Center for Integrated Quantum
  Science and Technologies, Universit\"at Ulm, D-89069 Ulm, Germany}
\newcommand{\saar}{Theoretische Physik, Universit\"at des Saarlandes, D-66123
  Saarbr\"ucken, Germany}
\newcommand{\padova}{Dipartimento di Fisica e Astronomia, Universit\`a degli
  Studi di Padova, I-35131 Italy}

\makeatletter
\def\l@subsubsection#1#2{}
\makeatother

\begin{document}

\title{One-dimensional many-body entangled open quantum systems
with tensor network methods}

\author{Daniel Jaschke}
\affiliation{\csm}
\author{Simone Montangero}
\affiliation{\ulm}
\affiliation{\saar}
\affiliation{\padova}
\author{Lincoln D.\ Carr}
\affiliation{\csm}


\begin{abstract}


  We present a collection of methods to simulate entangled dynamics of
  open quantum systems governed by the Lindblad master equation with tensor
  network methods.
  Tensor network methods using matrix product states have been proven very
  useful to simulate many-body quantum systems and have driven many innovations
  in research. Since the matrix product state design is tailored for closed
  one-dimensional systems governed by the Schr\"odinger equation, the next
  step for many-body quantum dynamics is the simulation of one-dimensional open
  quantum systems. We review the three dominant approaches to the simulation
  of open quantum
  systems via the Lindblad master equation: quantum trajectories, matrix
  product density operators, and locally purified tensor networks. Selected
  examples guide possible applications of the methods and serve moreover
  as a benchmark between the techniques. These examples include the finite
  temperature states of the transverse quantum Ising model, the dynamics of
  an exciton traveling
  under the influence of spontaneous emission and dephasing, and a double-well
  potential simulated with the Bose-Hubbard model including dephasing. We
  analyze which approach is favorable leading to the conclusion that a complete
  set of all three methods is most beneficial, pushing the limits of different scenarios.
  The convergence studies using analytical results for macroscopic variables
  and exact diagonalization methods as comparison, show, for example, that
  matrix product density operators are favorable for the exciton
  problem in our study.
  All three methods access the same library, i.e., the software
  package \emph{Open Source Matrix Product States}, allowing us to have a
  meaningful comparison between the different approaches based on the
  selected examples. For example, tensor operations are accessed from the
  same subroutines and with the same optimization eliminating one possible
  bias in a comparison of such numerical methods.

\end{abstract}

\maketitle

\setcounter{tocdepth}{1}
\tableofcontents

\section{Introduction                                                          \label{sec:otn:intro}}

The study of the combination of many-body quantum systems and open systems is one of the
critical pieces \emph{needed} to develop powerful quantum simulators and quantum computers.
While the many-body part is strictly necessary to scale these systems to
sizes useful for applications, open quantum systems help one to understand the
effects of decoherence and, therefore, the lifetime of the actual system.
Further questions are the analysis of steady states and the transient
dynamics approaching them. Reservoir engineering has the purpose of
preparing the system in a defined state; this state is equal to the steady
state and, therefore, protected from decoherence. The process of
thermalization is another example for open quantum system research.
The open system implementations in our \emph{Open Source Matrix Product States}
(\OSMPS{}) package combine two popular approaches: tensor networks for
many-body simulations and the Lindblad master equation as a default approach
for Markovian open quantum systems.

The Lindblad master equation~\cite{Kossakowski1972,Lindblad1976,Gorini1976}
is one common approach to open quantum systems~\cite{BreuerPetruccione,
Joye2006,Rivas2012,Schaller2014},
although its limitations are well-known. The benefit of this approach is the
conservation of the properties of the density matrix, i.e., norm and
positivity. The assumptions during the derivation, e.g., the Born-Markov and
secular approximation, limit the use of the Lindblad master equation to
quantum systems weakly coupled to large reservoirs. Other approaches to
open quantum systems are for example hierarchical equations of
motions~\cite{Tanimura1989,Tanimura1990} or various techniques for
non-Markovian open quantum systems~\cite{deVega2017}, but a versatile
implementation of the Lindblad equation is the first step towards the
implementation of a tensor network suite to study open quantum systems.

The history of tensor networks for quantum mechanics reaches back to the density matrix
renormalization group (DMRG) \cite{White1992,White1993,Schollwoeck2005}, then recasted into the
matrix product state (MPS) language \cite{Vidal2003,Schollwoeck2011,Orus2014}; recent
papers highlight the equivalence between the two approaches
\cite{Chan2016}. Based on the original MPS idea, many different
tensor networks have been derived ranging from tree-tensor
networks (TTN)~\cite{Shi2006} over the multi-scale entanglement
renormalization ansatz (MERA)~\cite{Vidal2008} to
projected entangled pair states (PEPS) \cite{Verstraete2008}, where the latter is
designed for two-dimesional systems; TTN and MERA can solve both
one-dimensional systems and generalizations to higher dimension.
Approaches such as PEPS can also be generalized to simulate open system
in two dimensions, as recently shown in \cite{Kshetrimayum2017}.
We focus in this work on one-dimensional systems in an MPS-like chain
structure. Tensor networks are very-well suited for this kind of
low-dimensional many-body system, where the area law describing the
scaling of the entanglement is the most favorable \cite{Eisert2010}. We
concentrate on tensor networks for the time evolution of open
systems and a background in MPS techniques is assumed. We emphasize
that the steady state of an open quantum system can be calculated
variationally \cite{Weimer2015,Cui2015,Mascarenhas2015}, which is not
included in this work. The point of this paper
is the side-by-side discussion of three different approaches relating
the technical implementation and their implications for the convergence
of actual simulations. The first of the three
approaches are quantum trajectories (QT)~\cite{Dalibard1992,Dum1992,
Wiseman1996,Plenio1998,Carmichael2009}
providing a statistical approach to the Lindblad master equation. In
contrast, both matrix product density operators
(MPDOs)~\cite{Verstraete2004,Zwolak2004} and locally purified tensor
networks (LPTNs)~\cite{Werner2016} simulate the complete density
matrix. The latter two approaches can also simulate thermal states, which
are otherwise only accessible through METTS~\cite{Stoudenmire2010} or after
building a sufficient number of eigenstates. Both MPDO and LPTN
representations have their limitations as discussed in
References~\cite{DeLasCuevas2013,Kliesch2014,DeLasCuevas2016}. Very briefly,
the arguments against each approach are that MPDOs do not conserve positivity
and the corresponding check is an NP-hard problem. In contrast, there are
states which have a representation in terms of MPDOs and maintain translational
invariance, while LPTNs leak a similar representation for this set of states.
We point out that there are previous comparisons between two of the
methods~\cite{Bonnes2014}.

Possible applications arise in the fast-evolving fields of quantum
simulators and quantum computing experiments. Rydberg systems are
one promising platform for quantum simulators, and
reference \cite{Weimer2010} outlines their possible applications within the
framework of the Lindblad master equation. The Lindblad operators described
therein are quasi-local, meaning acting on a neighborhood of sites.
The treatment of superconducting qubits coupled to phonon modes is another
architecture \cite{Chu2017}, allowing one to couple the superconducting qubit to
other degrees of freedom in the systems. The Lindblad equation was used in
this context to simulate the lifetime of the phonon modes \cite{Chu2017S}.
Trapped ions system are considered to be quantum simulators for open systems
themselves as pointed out in references~\cite{Barreiro2011,Schindler2013}.
Although the primary focus is on simulating open quantum systems according to
Kraus operators, the application of numerical simulations to this scenario
seems very fruitful to us given the connection between Lindblad equation and
Kraus operators \cite{Cappellaro2012}. The review in \cite{Mueller2012} highlights
the quantum simulator applications of Rydberg systems and trapped ions.
Moreover, it lists examples for ultracold atoms systems in an open system
context. The atomic, molecular, and optical (AMO) platforms provide
another set of problems to be studied. Atomic bosons can heat due to the
interaction with the optical lattice \cite{Pichler2013}, and molecules have
even more degrees of freedom \cite{Carr2011} to be used within open quantum
systems. The different internal degrees of freedom, i.e., rovibrational and
motional degrees of freedom, can be used for encoding individual reservoirs
for each molecule; one degree of freedom acts as a system, while another degree
of freedom acts as a reservoir for the first \cite{Branderhorst2006}.
This incomplete list shows the possible application of open
quantum systems in the quantum simulator context. We argue that the
consideration of system-environment effects will be an even more intensive focus
of future research as decoherence times of experiments increase and
errors decrease.

The outline of the paper is as follows. Section~\ref{sec:otn:methods}
is a very brief review of methods used to simulate open quantum systems
with tensor network methods. Section~\ref{sec:otn:simulation}
provides the actual details of the implementation in the OSMPS
package. We follow the structure of earlier work \cite{JaschkeMPS}
connecting the different time evolution methods to the open system.
The setup of the simulations and their convergence is discussed in
Sec.~\ref{sec:otn:conv}, where this section contains the showcases of
applications. The finite temperature states are the first example in
Sec.~\ref{sec:otn:conv:finitet}, where the quantum Ising model is one
possibility to describe many two-level quantum systems.
Section~\ref{sec:otn:conv:oqsa} turns to the transport of an exciton, which
travels under the influence of the interaction with an environment. The last
example considers a double-well potential governed by the Bose-Hubbard model,
where the oscillation between left and right well are damped out in the open
quantum system, see Sec.~\ref{sec:otn:conv:oqsb}. We conclude
in Sec.~\ref{sec:otn:conclusion}. Appendix~\ref{sec:otn:conv:oqsc}
provides additional aspects of non-local Lindblad operator for the example
of a dissipative state preparation; Appendix~\ref{app:inftbh} provides
technical details on the bond dimension of finite temperature states with
symmetries.

\section{Theoretical Approaches to the Simulation of Open Quantum Systems       \label{sec:otn:methods}}

Before going into details of the numerical setup for the simulation of the
Lindblad master equation, it is worthwhile to keep in mind the existing alternative
approaches for the simulation of open quantum systems. Amongst all the
different techniques which have been outlined for open quantum systems
are stochastic methods~\cite{Biele2012}, Redfield master
equations~\cite{BreuerPetruccione},
or solving the full system. Some of these methods are within the reach
of tensor network methods, e.g., the Redfield master equation. Others, i.e.,
the simulation of the full system, can already be achieved with MPS methods
as long as system plus environment together are not too big. The Lindblad
master equation is the first choice among above list as it conserves norm, Hermiticity, and
positivity of the state. OSMPS uses the Lindblad equation
\begin{eqnarray}                                                                \label{eq:otn:lindblad}
  \dot{\rho} &=& \frac{\i}{\hbar} [\rho, H] + \sum_{\nu} L_{\nu} \rho L_{\nu}^{\dagger}
                 - \frac{1}{2} \{L_{\nu}^{\dagger} L_{\nu}, \rho \} \, ,
\end{eqnarray}
which describes the evolution of the density matrix $\rho$ under the
Hamiltonian $H$ and a set of Lindblad operators $L_{\nu}$. Thereafter,
$\hbar$ is set to one. As we are
treating many-body systems, the index $\nu$ can be a combined index
running itself over the different Lindblad operators $\mu$ and different
sites $k$ in the system, i.e., $\nu = (\mu, k)$. The Lindblad equation
includes the approximations explained as follows. Ideally, given
unlimited resources, we would simulate the Schr\"odinger equation for the
system $S$ of interest and its environment $E$
\begin{eqnarray}                                                                \label{eq:otn:schroedinger}
  \frac{\partial}{\partial t} \ket{\psi_{S+E}}
  = - \frac{\i}{\hbar} H_{S+E}(t) \ket{\psi_{S+E}} \, .
\end{eqnarray}
Considering that the environment can be enormous and the Hilbert space grows
exponentially with system size, we apply the following three
approximations to Eq.~\eqref{eq:otn:schroedinger} to obtain the Lindblad
equation. (i)~System and environment are in a product state at $t=0$ and
stay in a product state over the time evolution, i.e., $\rho_{S+R}(t)
= \rho_{S}(t) \otimes \rho_{E}$. Correlations decay fast if the reservoir
is large and the reservoir remains unperturbed by the interaction with
the system. This assumption settles the timescales between the
environment $\tau_E$ and the system $\tau_S$: $\tau_E \ll \tau_S$.
(ii)~Furthermore, the timescale of the system holds $\tau_S \ll
\tau_{S, \mathrm{eq}}$; the equilibration time of the system
$\tau_{S, \mathrm{eq}}$ is longer
than the time step. (iii)~We truncate fast oscillating terms similar
to the rotating wave approximation when considering transitions of different
frequencies in the system, where their difference sets the timescale. These
approximations are formally described in terms of the Born-Markov
approximation, i.e., (i) and (ii), and the secular approximation, see (iii).

In order to simulate a system according to the Lindblad master equation, we
distinguish two paths. QTs evolve pure states sampling over a variety of
trajectories. This approach is motivated by the fact that in an experimental
setup every measurement projects the density matrix into a pure state; we
assume that the measurement outcomes are non-degenerate for each state for simplicity.
The QT approach models the probability for the projection into a specific
state. Therefore, a single simulation only reflects one possible outcome
of an experiment. To obtain the outcome for Eq.~\eqref{eq:otn:lindblad}, sampling over
different trajectories is necessary. One advantage of this method is that we
use pure states and there is no significant increase in the computational
scaling with
respect to an MPS simulation for each trajectory. The local dimension of the
MPS used in each QT is the same as for the MPS in a closed system. Furthermore,
any MPS can be used as initial state without increasing the bond dimension prior
to the time evolution. During the time evolution, we rely on the capabilities of
the MPS compression scheme to reduce entanglement, which is obsolete after a
quantum jump. The additional steps for choosing the Lindblad
operator to be applied to the MPS are not significant. The computational scaling
of the open quantum system is reflected in the number of trajectories. The
different trajectories can be easily parallelized across different cores
with MPI (Message Passing Interface).

On the other hand, $\rho$ can be directly simulated, e.g, mapping the density
matrix $\rho$ to a superket vector $\Lket{\rho}$
\cite{Verstraete2004,Zwolak2004} resulting in a Schr\"odinger-like equation.
The superket $\Lket{\rho}$ is constructed by building a vector out of all
entries in the density matrix $\rho$. The governing equation is then 
\begin{eqnarray}                                                                \label{eq:otn:lindblad:superket}
  \frac{\partial}{\partial t} \Lket{\rho} = \mathcal{L}(t) \Lket{\rho} \, ,
\end{eqnarray}
where the non-Hermitian Liouville operator $\mathcal{L}(t)$ corresponds to
the Hamiltonian and is defined as:
\begin{eqnarray}                                                                \label{eq:otn:lindblad:superketlong}
  \mathcal{L}(t) &=& - \frac{\i}{\hbar} H(t) \otimes \1
                 + \frac{\i}{\hbar} \1 \otimes H^T(t)                     \\
                + && \sum_{\nu} L_\nu \otimes (L_\nu^{\dagger})^T
                 - \frac{1}{2} \left( L_\nu^{\dagger} L_\nu \otimes \1
                 + \1 \otimes \left( L_\nu^{\dagger} L_\nu \right)^T \right) \, . \nonumber
\end{eqnarray}
Overall, this approach allows us to simulate systems replacing the Hamiltonian
with $\mathcal{L}$ at the cost of increased local dimension. If the total
dimension of the closed system is $D$, the dimension of the problem in Liouville
space is $D^2$.

Finally, we can evolve the purification of $\rho$ denoted with $X$
\cite{Werner2016}. Since the density matrix is by definition positive, we
can decompose it into
\begin{eqnarray}                                                                \label{eq:otn:purification}
  \rho = X X^{\dagger} \, .
\end{eqnarray}
For example, a pure state $\ket{\psi}$ is equal to its purification
$X = \ket{\psi}$. To preserve the structure of the complex
conjugate pair, we use instead of Eq.~\eqref{eq:otn:lindblad} the more general Kraus
operators representing a completely positive trace preserving map (CPT map):
\begin{eqnarray}                                                                \label{eq:otn:kraus}
  \rho(t + dt) = \sum_{\nu'} K_{\nu'} \rho(t) K_{\nu'}^{\dagger}
\end{eqnarray}
$K_{\nu'}$ are called Kraus operators. There is a connection to derive the
Kraus operators via Choi's theorem from the Lindblad equation~\cite{Choi1975},
or we approximate the Kraus operators in first order in $dt$ and truncating
higher order terms of $dt$ when expanding $\rho(t + dt) \approx \rho(t)
+ dt \dot{\rho}(t)$ with $\dot{\rho}(t)$ as defined in the Lindblad master
equation~\eqref{eq:otn:lindblad}~\cite{Cappellaro2012}.

The MPS algorithms profit from considerable speed-ups when
symmetries are present in the system and encoded to the tensor network. We
distinguish between Abelian symmetries, considered in the following, and
non-Abelian symmetries. Abelian groups are preferable from the perspective of
an implementation as operations commute
and keeping track of quantum numbers reduces to a simple group operations; in
contrast, non-Abelian groups have to use the Clebsch-Gordan coefficients.
We can use symmetries in the Schr\"odinger equation when the commutation of
some operator $G$ commutes with the Hamiltonian, $[H, G] = 0$, and has a
definition in terms of the local Hilbert spaces. For example, the Bose-Hubbard
model conserves the number of particles and $N = \sum_{k} n_{k}$ commutes with the
Bose-Hubbard Hamiltonian, see later on in Eq.~\eqref{eq:otn:ham:bosehubbard}.
We can also use the Abelian symmetry in the Liouville equation if the Lindblad
operators do not violate the symmetry,
\begin{eqnarray}                                                                \label{eq:otn:liouvillesymm}
  \left[ \mathcal{L}, G \otimes \1 + \1 \otimes G^{T} \right] = 0 \, .
\end{eqnarray}
Thus, we cannot add loss of particles with an annihilation operator as
a Lindblad operator in a number-conserving simulations: the Lindblad
operator violates the symmetry.

This brief overview enables us to take a closer look at the different
evolution methods in Sec.~\ref{sec:otn:simulation}.

\section{Tensor Networks Simulations of the Lindblad Master Equation                             \label{sec:otn:simulation}}

This section covers the technical aspects of all three approaches to open quantum
systems. We start with the simulation of the Lindblad master equation in terms
of MPDOs\cite{Verstraete2004,Zwolak2004} and describe in detail the issues
raised by a non-Hermitian operator. The quantum
trajectories~\cite{Dalibard1992,Dum1992,Daley2014} share the non-Hermitian operator with
the MPDOs, and we explain quantum trajectories in the following section.
Finally, we discuss the third approach using LPTNs~\cite{Werner2016}.
Once we have covered these aspects, we turn to the convergence of the different
approaches. Readers aware of the technical details or solely interested in
the practical application of the methods may, therefore, skip this section
and to straight to Sec.~\ref{sec:otn:conv}.

\subsection{Matrix Product Density Operators                                   \label{sec:otn:mpdo}}

The formulation of MPDOs heavily relies on the Liouville operator and superket
notation in Eq.~\eqref{eq:otn:lindblad:superketlong} and the similarity to the
Schr\"odinger equation. The first steps
into the MPDOs implementation inside \OSMPS{} describe the notations of the
superket $\Lket{\rho}$ and we first introduce the static aspect of MPDOs,
i.e., their ability to calculate thermal states. Then, we move toward
matrix product operators (MPOs) \cite{Pirvu2010,WallNJP2012} of the Liouville operator.
Knowing these two principal objects, i.e., the superket and the MPO, we move
forward to the time-evolution methods using the MPO, i.e., Krylov, local
Runge-Kutta (LRK) \cite{Zaletel2015}, and the time-dependent variational
principle (TDVP) \cite{Haegeman2016}. Finally, we fill in the description of the Krylov
time-evolving block decimation (KTEBD) and the well-established time-evolving
block decimation (TEBD) \cite{Vidal2003} algorithm and its modifications for the evolution
of the Lindblad master equation. In fact, the time evolution methods for
the quantum trajectories revisit many issues already discussed here as
both methods deal with a non-Hermitian propagator.

\subsubsection{Construction of Matrix Product Density Operator states          \label{sec:otn:mpdo:states}}

We limit ourselves to the construction of two classes of initial states for
the time evolution. On the one hand, we convert MPS states to MPDOs.
Suitable MPS states include product states, ground states, low-lying excited
states, and pure states obtained via unitary time evolution.
On the other hand, we would like to have finite temperature states via an
imaginary time evolution of MPDOs. Therefore, the infinite temperature state
has to be given as an initial state for the imaginary time evolution as it
serves as a starting point for cooling. This procedure is explained in
Sec.~\ref{sec:otn:mpdo:imagt}. With the knowledge of the infinite
temperature state which is a mixed product state, the construction for
product states of any other type can be derived.

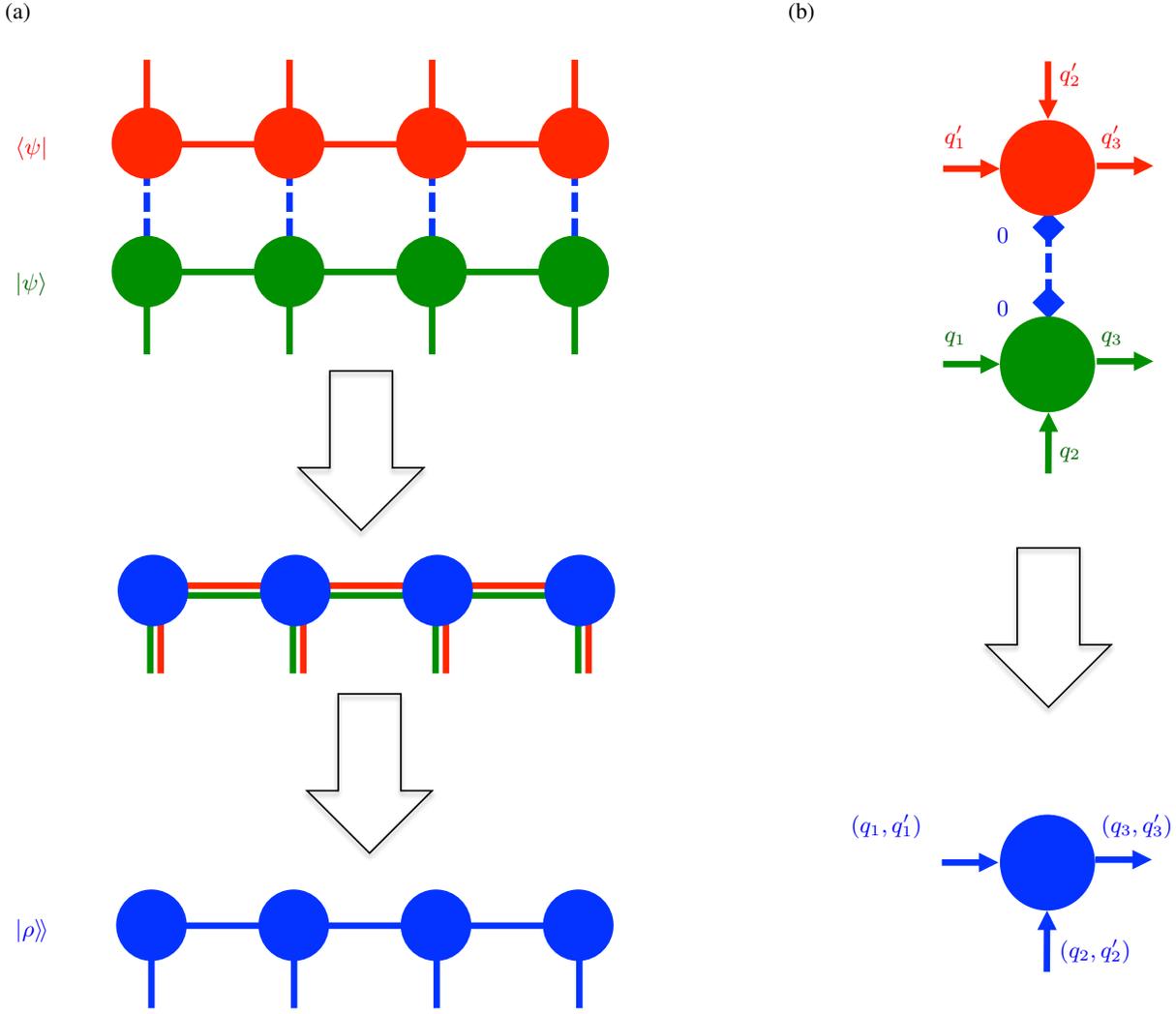
\begin{figure}[t]
 \begin{center}
   \vspace{0.7cm}
   \begin{overpic}[width=0.8 \columnwidth,unit=1mm]{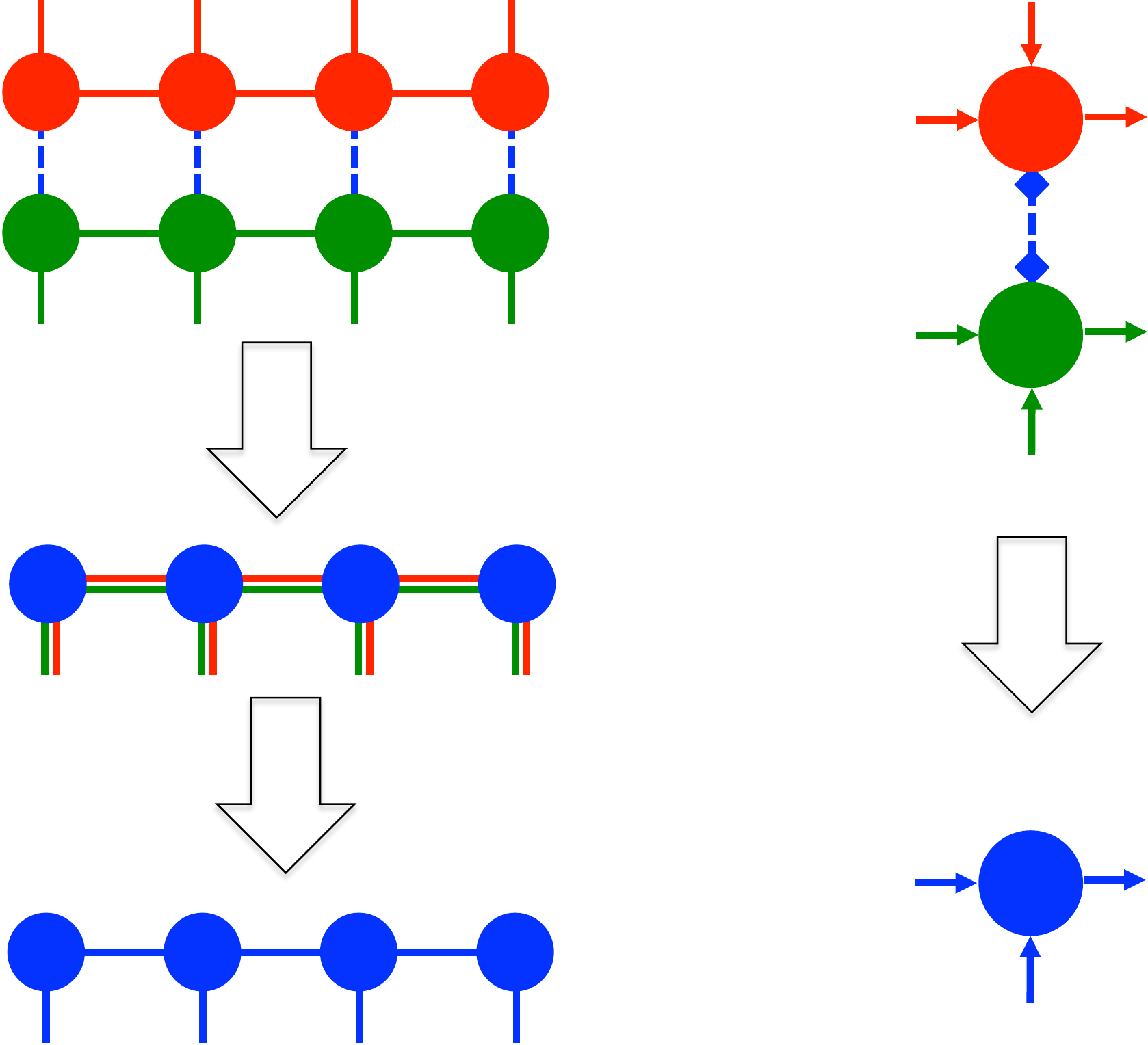}
     \put(-10, 95){(a)}
     \put(65, 95){(b)}
     \put(-9, 82){\color{red} $\bra{\psi}$}
     \put(-9, 69){\color{dg} $\ket{\psi}$}
     \put(-9, 7){\color{blue} $\Lket{\rho}$}
     \put(91, 89){\color{red}$q_2'$}
     \put(80, 83){\color{red}$q_1'$}
     \put(95, 83){\color{red}$q_3'$}
     \put(95, 64){\color{dg}$q_3$}
     \put(80, 64){\color{dg}$q_1$}
     \put(91, 53){\color{dg}$q_2$}
     \put(85, 66.5){\color{blue}$0$}
     \put(85, 73.5){\color{blue}$0$}
     \put(71, 17){\color{blue}$(q_1, q_1')$}
     \put(91, 5){\color{blue}$(q_2, q_2')$}
     \put(95, 17){\color{blue}$(q_3, q_3')$}
   \end{overpic}
   \caption[MPDO from MPS state]
   {\emph{MPDO from MPS state.} (a)~In an MPS without symmetries,
   each site tensor is contracted over a one-dimensional auxiliary link
   with its complex conjugated counterpart. The corresponding links are
   fused. (b)~The auxiliary tensor contracted over in an MPS with symmetries
   is labeled with a dummy quantum number 0. We get all combination of
   sub-tensors. In addition to merging the links of the sub-tensors, the
   quantum numbers are combined.
                                                                                \label{fig:otn:mps2mpdo}}
 \end{center}
\end{figure}

Figure~\ref{fig:otn:mps2mpdo} describes the transformation of an MPS into an
MPDO. If the tensor $A_{\alpha, i, \beta}^{[k]}$ represents the site $k$ in
the MPS, we add an auxiliary link of dimension one and obtain
$A_{\alpha, i, \kappa, \beta}^{[k]}$. We contract this tensor with its
complex conjugate tensor $\left(A_{\alpha', i', \kappa, \beta'}^{[k]}
\right)^{\ast}$ over $\kappa$ leading to the MPDO representation
$B_{(\alpha, \alpha'), (i, i'), (\beta, \beta')}$. The contraction over
the auxiliary index corresponds to the outer product $\ket{\psi} \bra{\psi}$,
taken locally on site $k$. Thus, the usage of an outer product is an
alternative to the auxiliary index. The indices in parentheses represent
fused indices; therefore, the new tensor is again of rank 3. The fusion
of two indices produces a new index using a Cartesian product to map the
elements, e.g., $\alpha, \alpha' \to (\alpha, \alpha') = \alpha''$. The
dimension of the new index $\alpha''$ is the product of the dimension
of the two original indices $\alpha$ and $\alpha'$. We notice the
increase in the dimension of links: for a tensor with bond dimensions
$(\chi, d, \chi)$ in the MPS, the new bond dimensions in the MPDO are
$(\chi^2, d^2, \chi^2)$. The number of sub-tensors for a symmetric tensor
network also increases. If we have $n$ sub-tensors in an MPS site, the
MPDO representation of the site has $n^2$ sub-tensors.

The infinite temperature state $\rho_{\infty}$ is straightforward to implement in an
MPDO without symmetries but has some subtleties when symmetries are used.
The density matrix $\rho_{\infty}$ is the identity matrix normalized to trace one. If we
do not have symmetries, the global identity is a product of local identities
$\rho_{\infty} \propto \bigotimes_{k=1}^{L} 1_{k}$. Thus, the bond dimension
between sites in the MPDO is $\chi = 1$, and the tensors are
$B_{\alpha, (i, i'), \beta} = \delta_{ii'}$. In contrast, the product of local
identity for a symmetric MPDO contains states in all possible symmetry
sectors and does not lead to $\rho_{\infty}$ for a specific sector. Instead, we
construct the state $\rho_{\infty}$ representing the Gibbs distribution at
infinite temperature from symmetric sub-tensors on each site $k$ filled with
\begin{eqnarray}
  B_{\alpha, (i, i'), \beta}^{[k]}((q_1, q_1'), (q_2, q_2'), (q_3, q_3'))
  = \delta_{i i'} \delta_{q_1 q_1'} \delta_{q_2 q_2'} \, . \nonumber \\
\end{eqnarray}
The incoming quantum number from the left must be chosen in such a way that
is possible to obtain it with the local Hilbert spaces on the left.
The outgoing quantum number $q_3 = q_1 + q_2$ must allow us to have the
sector specified: $q_3^{[k=L]}$ on the last site must match the global sector exactly.
We illustrate these choices with the Bose-Hubbard model with a
maximum filling of $3$, i.e., local dimension $d = 4$. Further, we specify
the system size $L = 5$ and unit filling. $q_1^{[k=2]} \in \{0, 1, 2, 3 \}$ for the second site
$k = 2$, $q_1^{[l=2]} = 4$ cannot be achieved from a single site on the left.
Similar, for the fourth site $k = 4$, $q_3^{[k=4]} \in \{ 2, 3, 4, 5 \}$. Any lower
filling cannot reach unit filling with a single site on the right. Higher
$q_3^{[k=4]}$ are already above unit filling and do not match the symmetry
sector. In conclusion, the symmetric $\rho_{\infty}$ has already a bond
dimension $\chi > 1$ from the beginning and can contain a large number
of sub-tensors. For example, the bond dimension of the Bose-Hubbard model
on $L$ sites with maximal filling $d - 1$ for each site has an upper bound for the bond
dimension of $d (N - d + 2)$ distributed on the block-diagonal structure. We
present a detailed calculation in Appendix~\ref{app:inftbh}. A generalization
to more complicated symmetries is possible, but a generalized equation for the
maximal bond dimension of any model and symmetry is difficult to obtain.

\subsubsection{Imaginary time evolution for finite-T states                    \label{sec:otn:mpdo:imagt}}

The finite temperature (finite-T) states are based on the Gibbs distribution
defined as $\rho(T) = \exp(- H / (k_B T)) / \mathcal{N}$ with $\mathcal{N}
= \mathrm{Tr} \left[\exp(- H / (k_B T)) \right]$ and $k_B = 1$ the Boltzmann
constant and have been pointed
out as a feature of MPDOs from their definition on~\cite{Verstraete2004,
Zwolak2004}. From the Gibbs distribution, we rewrite the unnormalized state
as
\begin{eqnarray}
  \mathrm{e}^{- H / (k_B T)}
  &=& \mathrm{e}^{- H / (2 k_B T)} \1 \mathrm{e}^{- H / (2 k_B T)} \nonumber \\
  &\propto&  \mathrm{e}^{- H / (2 k_B T)} \rho_{\infty} \mathrm{e}^{- H / (2 k_B T)} \nonumber \\
  &=& \mathrm{e}^{\frac{1}{2 k_B T} \left(- H \otimes \1 - \1 \otimes H^T \right)} \Lket{\rho_{\infty}} \, ,
\end{eqnarray}
which corresponds to an imaginary time evolution similar to the ground state
algorithm of an MPS. The operator in Liouville space is Hermitian. In
detail, we enable the TEBD algorithms for imaginary time evolution. The
definition of the initial state $\Lket{\rho_{\infty}}$ was discussed in the
previous Sec.~\ref{sec:otn:mpdo:states} as an example of how to construct
initial states represented as an MPDO.

\subsubsection{Matrix Product Operators in Liouville space                     \label{sec:otn:mpdo:mpo}}

For the simulation of the Lindblad master equation with matrix MPO techniques, we
transform Eq.~\eqref{eq:otn:lindblad} into the Liouville space. This transformation allows us
to represent the density matrix as vector $\Lket{\rho}$ and use MPS
techniques developed for pure states on an enlarged local space. The local
dimension $d$ for pure states $\ket{\psi}$ translates into a local dimension
$d_{\mathcal{L}} = d^2$ for the simulation of $\Lket{\rho}$.
In recent work~\cite{JaschkeMPS}, we presented the evolution techniques in the \OSMPS{} package
for pure states based on formulating the Hamiltonian $H$ of the system as an MPO.
We can reuse all evolution techniques for MPDOs once we can formulate
$\mathcal{L}$ as an MPO. We restrict the Hamiltonian to local terms and bond terms,
i.e., nearest neighbor interactions, to explain the procedure.
But the methods are generalized to any rule set present in \OSMPS{} for
the Liouville space.
For the MPS evolution, we know that the bond dimension of the MPO for $n_{\mathrm{site}}$
local terms and $n_{\mathrm{bond}}$ nearest neighbor terms is $\chi_{\mathrm{MPO}} = 2 + n_{\mathrm{bond}}$.
We explain how to build the MPO matrices along the quantum Ising model with the Hamiltonian
\begin{eqnarray}                                                                \label{eq:otn:HQI}
  H_{\mathrm{QI}} &=& -J \sum_{j=1}^{L-1} \sigma_{j}^{z} \sigma_{j+1}^{z}
                      - g \sum_{j=1}^{L} \sigma_{j}^{x} \, ,
\end{eqnarray}
where $\sigma_{j}^{x}$ and $\sigma_{j}^{z}$ are the Pauli matrices acting
on site $j$. The interaction strength between neighboring spins is $J$, the
coupling to the external field is $g$, and the system size is $L$.
The MPO operator-valued matrix in the bulk of the system for the Hamiltonian
$H_{\mathrm{QI}}$ is
\begin{eqnarray}
  M_{H} &=& \begin{pmatrix}
    \1            &0               &0 \\
    \sigma^{z}    &0               &0 \\
    -g \sigma^{x} & - J \sigma^{z}  & \1
  \end{pmatrix} \, ,
\end{eqnarray}
where the two of the four dimensions of the rank 4 MPO tensor are
encoded into the row and columns of this matrix. These two indices are
contracted with left and right neighboring MPO-matrices and the number
of rows/columns is the bond dimension of the MPO. The first (last) MPO
is an operator-valued row (column) vector. The remaining two
dimensions are the dimensions of the matrix for each entry. In this notation, $0$ is to
be understood as a $d \times d$ matrix of zeros. A convenient
implementation uses operator-valued sparse matrices instead of building
the full rank 4 tensor with a large fraction of zeros~\cite{WallNJP2012}.
We transform this MPO to the Liouville space according to
Eq.~\eqref{eq:otn:lindblad:superket}; the relevant terms are the Hamiltonian
terms $H \otimes \1 - \1 \otimes H^{T}$. We reorder the subspaces such that
the terms for each site are collected. The two links for the Hilbert space
of each site have been fused previously. Thus, the new operator-valued
entries have a dimension of $d^2 \times d^2$. This transformation increases
the MPO bond dimension to $\chi_{\mathrm{MPO}, \mathcal{L}} = 2 + 2 n_{\mathrm{bond}}$:
%
\begin{eqnarray}                                                                 \label{eq:otn:HMPOLiou}
  M_{\mathcal{L}} &=&
  \begin{pmatrix}
    \1                          &0               &0               & 0 \\
    \sigma^{z} \otimes \1        &0               &0              & 0 \\
    \1 \otimes (\sigma^{z})^T    &0               &0              & 0 \\
    -g \sigma^{x} \otimes \1 - \1 \otimes -g (\sigma^{x})^T
    & - J \sigma^{z} \otimes 1
    & - (\1 \otimes - J (\sigma^{z})^{T}) & \1
  \end{pmatrix} \, . \nonumber \\
\end{eqnarray}
Here in the construction of $M_{\mathcal{L}}$, we assume that the
$\mathcal{L}$ is multiplied with the usual
$(- \i \dt)$ in the evolution. The different signs in front of
$H \otimes \1$ and $\1 \otimes H^{T}$ are considered together with the
coupling constant of the Hamiltonian term and are not double-counted.
So far, this step allows us to evolve a closed
initially pure or mixed system. To include local Lindblad
operators $L_{\mu}$ acting on each site as part of the dissipative part
of the evolution, we add the following matrix based on
Eq.~\eqref{eq:otn:lindblad:superket} to the previous result in
Eq.~\eqref{eq:otn:HMPOLiou}, which contains the Hamiltonian part of the
evolution in Liouville space:
%
\begin{eqnarray}
  M_{\mathcal{L}} &=&
  \begin{pmatrix}
    0 & 0 & 0 & 0 \\
    0 & 0 & 0 & 0 \\
    0 & 0 & 0 & 0 \\
    \i \gamma \left(L_{\mu} \otimes L_{\mu}^{\ast} - \frac{1}{2} L_{\mu}^{\dagger} L_{\mu} \otimes \1
                      - \frac{1}{2} \1 \otimes L_{\mu}^{T} L_{\mu}^{\ast} \right)
    & 0 & 0 & 0
  \end{pmatrix} \, . \nonumber \\
\end{eqnarray}
The imaginary unit takes into account the multiplication of $(-\i \dt)$
leaving only $\dt$ after the construction of the argument for the
exponential. Thus, the local Lindblad operators do not increase the
bond dimension of the MPO. We implement another type
of Lindblad operator similar to a Hamiltonian many-body string term,
i.e., the many-body string Lindblad operator
\begin{eqnarray}
  L_k(r) = L_k \otimes L_{k+1} \otimes \cdots \otimes L_{k+r-1} \, .
\end{eqnarray}
The MPO bond dimension for a many-body string Hamiltonian term is
$r - 1$ for the Hamiltonian, and $2 (r - 1)$ for the Liouville operator. The many-body string
Lindblad term has a bond dimension of $3 (r - 1)$. In detail, we implement
three terms with bond dimension $(r - 1)$, which are the three different
terms in the dissipative part
\begin{subequations}\begin{eqnarray}                                            \label{eq:otn:MPOsuba}
  && L_k \otimes \cdots \otimes L_{k+r-1} \, \rho \,
  L_k^{\dagger} \otimes \cdots \otimes
  L_{k+r-1}^{\dagger} \, , \\                                                   \label{eq:otn:MPOsubb}
  && - \frac{1}{2} L_k^{\dagger} L_k \otimes \cdots \otimes
  L_{k +  r - 1}^{\dagger} L_{k + r - 1} \, \rho \, , \\                        \label{eq:otn:MPOsubc}
  && - \frac{1}{2} \rho \, L_k^{\dagger} L_k \otimes \cdots \otimes
  L_{k +  r - 1}^{\dagger} L_{k + r - 1} \, .
\end{eqnarray}\end{subequations}
The tripling of the bond dimension for a multi-site Lindblad rule
equivalent to a Hamiltonian rule holds in general.
Equation~\eqref{eq:otn:MPOsuba} represents $L \rho L^{\dagger}$
in the Lindblad equation, see Eq.~\eqref{eq:otn:lindblad}. The other
two terms, i.e., Eqs.~\eqref{eq:otn:MPOsubb} and \eqref{eq:otn:MPOsubc},
build the anti-commutator.
The setup of $\mathcal{L}$ as an MPO is the primary step to use
the evolution methods. Nonetheless, the MPS methods cannot be used blindly
since $\mathcal{L}$ does not necessarily maintain Hermiticity in contrast to the
Hamiltonian, i.e., $\mathcal{L}^{\dagger} \neq \mathcal{L}$. The Hamiltonian
contributions in Eq.~\eqref{eq:otn:lindblad:superket} are still Hermitian, factoring
out the imaginary unit $\i$. $ L_{\nu}^{\dagger} L_{\nu} \otimes \1$  and $\1 \otimes
(L_{\nu}^{\dagger} L_{\nu})^{T}$ are Hermitian themselves, but not with the
additional $-\i$ factored out from the Hamiltonian. $L_{v} \otimes
(L_{\nu}^{\dagger})^{T}$ is not necessarily Hermitian in itself or with an
additional $-\i$.
In the following, we briefly discuss the adaptions for each method with regards
to the closed system MPS implementation.

\subsubsection{Krylov-Arnoldi subspace method                                  \label{sec:otn:mpdo:krylov}}

The Krylov approximation~\cite{Moler2003,Schmitteckert2004,Garcia2006time,
WallNJP2012} directly builds the new state after
the time step $\dt$ evaluating the product $\exp(\mathcal{L} \dt)
\ket{\psi}$. The exponential taken within the Krylov subspace is much
smaller than the dimension of the Hilbert space or Liouville space
scaling with $d^{L}$ and $d^{2L}$, respectively. The exponential in the
Krylov subspace is used to approximate the new state after the time step;
thus, the exponential does not represent a propagator.
We recall that in case of a Hamiltonian, which is by definition
Hermitian, the matrix to be exponentiated in Krylov subspace is tridiagonal.
We label it Krylov-Lanczos in analogy to the Lanczos eigenvalue algorithm for
Hermitian matrices. In contrast, the Liouville operator may violate Hermiticity,
and we use the Krylov-Arnoldi algorithm. We use the name Krylov-Arnoldi
due to the similarity with the Arnoldi algorithm solving for eigenvalues of
a non-Hermitian matrix. Both variants of the algorithm construct a set
of orthogonal Krylov vectors $\{ \vec{v}_{\eta} \}, \; \eta = 0, \ldots, M$
from the powers of the operator, i.e., Hamiltonian or Liouville operator.
The number of Krylov vectors $M$ is determined based on a tolerance and
much smaller than the complete space, e.g., $d^{2L}$ for the Lindblad
master equation. The initialization of the Krylov vectors and the iterative
construction follows

\begin{eqnarray}
  \vec{v}_{0} &=& \ket{\psi} \, , \qquad
  \vec{v}_{0} = \frac{\Lket{\rho}}{\Lbra{\rho}\Lket{\rho}} \, , \\
  \vec{v}_{\eta + 1}' &=& H \vec{v}_{\eta} \, , \qquad
  \vec{v}_{\eta + 1}' = \mathcal{L} \vec{v}_{\eta} \, ,
\end{eqnarray}
where $\vec{v}_{\eta}'$ have to be orthogonalized against the previous
Krylov vectors to obtain $\vec{v}_{\eta}$. Taking a step back from the
details, one observes that this procedure is well-defined in terms of tensor
networks. The application of the Hamiltonian or Liouville operator
represented as an MPO to a quantum state can be either executed via a
contraction followed by a compression or fitting a state while minimizing
the distance. The orthogonalization depends of the overlap of two vectors
represented as an MPS or MPDO, which is a standard implementation for
measuring distances between pure states. In a second step, the vectors
weighted with the overlap have to subtracted from $\vec{v}_{\eta}'$.
Sums of MPSs or MPDOs can be calculated variationally minimizing the
distance again; an actual sum increases the bond dimension and requires
compression becoming inconvenient beyond a few terms in a sum. The
variational methods to achieve these steps are explained for example
in \cite{WallNJP2012}.

So far, we have constructed the Krylov vectors but not propagated the
quantum state, MPS or MPDO, for one time step $dt$. A detailed description
of the construction of the Krylov matrix $M_{K}$ and proof of validity is
beyond the scope of this work, and we refer to the corresponding literature
\cite{Gallopoulos1992,Saad1992,Moler2003}. The Krylov matrix is constructed
from the overlaps of $\vec{v}_{\eta}$ and is sparse, i.e., tridiagonal (upper
Hessenberg) for a Hermitian Hamiltonian (Liouville operator). We introduce
the exponential
\begin{eqnarray}                                                                \label{eq:otn:krylprop}
  P_{K} &=& \mathrm{e}^{- \mathrm{i} M_{K} dt} \, , \qquad
  P_{K} = \mathrm{e}^{M_{K} dt} \, .
\end{eqnarray}
The $i^{\mathrm{th}}$ row and $j^{\mathrm{th}}$ column is specified with
$(P_{K})_{i,j}$. The new state propagated from $t$ to $t + dt$ is then
defined as
\begin{eqnarray}                                                                \label{eq:otn:krylupdate}
  \ket{\psi} &=& \sum_{i=0}^{M} (P_{K})_{i,0} v_{i} \, , \qquad
  \Lket{\rho} = \sum_{i=0}^{M} (P_{K})_{i,0} v_{i} \, .
\end{eqnarray}
The two final numerical steps include the implementation of the
matrix exponential in Eq.~\eqref{eq:otn:krylprop} and the summation in
Eq.~\eqref{eq:otn:krylupdate} to build the new state. The first is solved
for the Lindblad master equation with a general matrix exponential handling
the upper Hessenberg matrix $M_{K}$; the Hamiltonian version can profit
from using the tridiagonal Hermitian structure in $M_{K}$ and reflects
the main difference between the implementation of the Lindblad master
equation in contrast to the Schr\"odinger equation. In theory, one can use the
Krylov-Lanczos algorithm as a fallback for Hamiltonian mixed state evolutions
when handling the imaginary unit $\mathrm{i}$ accurately. The update of the
state is a sum over MPS or MPDOs and can be solved as aforementioned in the
orthogonalization. In summary, the MPDO representation is convenient in
combination with representing the MPO in Liouville space; all methods
except the exponential of the Krylov matrix can be reused without further
modification, keeping an implementation cheap.

\subsubsection{Local Runge-Kutta                                               \label{sec:otn:mpdo:lrk}}

The local Runge-Kutta (LRK) method is another method allowing us to evolve
Hamiltonians with long-range interactions~\cite{Zaletel2015}. The version
for the Schr\"odinger equation takes the MPO of the Hamiltonian and
calculates an MPO representation of the propagator for the corresponding
time step. The propagator MPO has a smaller bond dimension, by one, and
is an efficient representation. It can either be contracted to
obtain the new state or fitted. The steps to obtain the MPO of the
propagator involve an intermediate mapping to hard-core bosons. A
generalization to the non-Hermitian Liouville operator including
non-local Lindblad operators is beyond this
work here. But we can use the specific structure of the MPO for the
propagator to generalize it at least to local Lindblad operators and
their representation in Liouville space.

Therefore, we look at the operator-valued MPO matrix of the
propagator $W^{II}$ consisting of operator-valued four sub-matrices
with subscript $A$, $B$, $C$, and $D$:
\begin{eqnarray}
  W^{II} &=& \begin{pmatrix} W_{D}^{II} & W_{C}^{II} \\
               W_{B}^{II} & W_{A}^{II} \end{pmatrix} \, ,
\end{eqnarray}
Every term expect $W_{D}^{II}$ involves the mapping to hard-core bosons;
therefore, we do not describe $W_{A}^{II}$, $W_{B}^{II}$, and $W_{C}^{II}$ as
their definition does not change for local Lindblads and details can be found
in \cite{Zaletel2015}. $W_{D}^{II}$ contains all local site terms and is
simply the exponential of these local terms, i.e., the propagator of a
system truncating all interactions. For example, this local site term is the
coupling to the transverse field in the quantum Ising model with a Pauli matrix
or the number operator $n_{k}$ acting on site $k$ and the on-site interaction
$n_{k} (n_{k} - 1)$ for the Bose-Hubbard model with appropriate weight for both
cases. In the implementation of the Schr\"odinger equation,
$D$ is the sum over all site rules for site $k$ and therefore Hermitian. The
corresponding exponential uses this fact. If the Lindblad operators are local,
the Lindblad operators can be included entirely in this term $D$ in the MPDO
approach. The exponential has then to be calculated for a general matrix
because the Lindblad terms $L_{\nu} \otimes L_{\nu}^{\ast}$ do not necessarily
enforce Hermiticity. Since the other terms are not affected by the local terms,
they can stay in place as they are.

Evidently, this approach only works for local Lindblad operators. To what extent non-local
Lindblad operators are covered by the method remains a subject of future
research. Moreover, the current implementation makes the symmetric tensor
to the full space and back to calculate the representation for the
LRK-propagators. For that reason, the symmetry implementation only needs
to consider a correct mapping.

\subsubsection{Time-Dependent Variational Principle                            \label{sec:otn:mpdo:tdvp}}

The TDVP~\cite{Haegeman2011,Haegeman2016} is the third evolution method
based on the MPO. Its elegance is the elimination of errors depending on
the time step $dt$ for time-independent Hamiltonians; remaining errors are an
insufficient bond dimension or discretizing a time-dependent Hamiltonian in
time. The algorithm itself benefits from approaches used widely in other
tensor network algorithms, namely effective Hamiltonians from a variational
ground state search and similarities to the KTEBD algorithm. The differences
from a Hamiltonian evolution and the Liouville operator have to be
considered for its adaption. We point out that there is another
time evolution using directly a variational approach which minimizes the
distance between a guess for the new density matrix and the time-evolved
density matrix \cite{Weimer2016}.

Based on the suggested TDVP algorithms with a single-site update or a
two-site update, we use for our analysis the two-site version. The
advantages are the possibility for a growing bond dimension and the
automatic introduction of new symmetry sectors, which may not be present
in the initial state. The latter is, for example, important if the initial
state is defined as a product state, e.g., Fock state in the Bose-Hubbard
model. Reference \cite{Haegeman2016} derives in detail that the time
evolution is then defined in terms of the time evolution under effective
two-site operators and a backward time evolution of the single-site
operators. For the Hamiltonian version, one has to be able to calculate
\begin{eqnarray}
  \ket{\psi'} = \mathrm{e}^{- \mathrm{i} H_{\mathrm{eff}}^{[k,k+1]} dt} \ket{\psi} \, , \qquad
  \ket{\psi''} = \mathrm{e}^{+ \mathrm{i} H_{\mathrm{eff}}^{[k+1]} dt} \ket{\psi'} \, .
\end{eqnarray}
Both can be efficiently computed, i.e., even without involving
the variational methods used in Sec.~\ref{sec:otn:mpdo:krylov}, if the
orthogonality center is contained in the sites $(k, k+1)$ for the two-site
version and in $k+1$ for the single-site update; the corresponding tensors
form a vector space. The corresponding term $H_{\mathrm{eff}} \ket{\psi}$
can be calculated; thus, we can use the Krylov subspace method relying on
powers $H_{\mathrm{eff}}^{n} \ket{\psi}$ and update the corresponding
two-site tensor for $(k, k+1)$ (one-site tensor for $k + 1$) in the time
evolution step (backward evolution). The TDVP method does not make use
of the Hermitian property of the Hamiltonian, but requires the matrix
exponential. The Krylov-Lanczos method with the tridiagonal matrix is
used to calculate the propagator in the closed system case. The MPDO
profits from its close similarity to the MPS. The Liouville operator
is already represented as an MPO and $\mathcal{L}_{\mathrm{eff}}^{[k,k+1]}$
and $\mathcal{L}_{\mathrm{eff}}^{[k]}$ are constructed without any
adaption. In contrast to the Hamiltonian in the Schr\"odinger equation,
we have to take into account that $\mathcal{L}$ is not hermitian; thus,
the algorithm is again adapted for the Hessenberg matrix
yielded by the Krylov-Arnoldi algorithm, and it is implemented in OSMPS.
Section~\ref{sec:otn:mpdo:krylov} explains the differences between the
evolution under a Hermitian operator and a non-Hermitian operator with the
Krylov method, which are independent of using the method with a global
operator as in Sec.~\ref{sec:otn:mpdo:krylov} or with an effective, local
version as in the TDVP adaption.
As an outlook, this upper Hessenberg matrix reappears when using the TDVP with
the non-hermitian ``effective'' Hamiltonian of the quantum trajectories, where
``effective'' corresponds to the inclusion of dissipate terms and is further
broken down into the ``effective'' Hamiltonian acting on two sites.

\subsubsection{Time-Evolving Block Decimation                                  \label{sec:otn:mpdo:tebd}}

The TEBD \cite{Vidal2003} approach approximates the global propagator with
local propagators; most implementation target nearest-neighbor Hamiltonians
and the local propagators act on two sites. The global propagator
$\exp\left(\mathcal{L} dt \right)$ can use this technique. The
Suzuki-Trotter decomposition \cite{NielsenChuang} then approximates the
exponential of the Liouville operator as
\begin{eqnarray}                                                                \label{eq:otn:Ltrotter}
  \mathrm{e}^{\mathcal{L} dt} &=&
  \mathrm{e}^{\sum_{k=1}^{L/2} \mathcal{L}_{2k-1,2k} \frac{dt}{2}}
  \mathrm{e}^{\sum_{k=1}^{L/2-1} \mathcal{L}_{2k,2k+1} dt}
  \mathrm{e}^{\sum_{k=1}^{L/2} \mathcal{L}_{2k-1,2k} \frac{dt}{2}}
  + \mathcal{O}(dt^3)
\end{eqnarray}
where the formula in Eq.~\eqref{eq:otn:Ltrotter} represents the second
order decomposition for an even system size; upper bounds of the sum
over the site index $k$ have to be adapted for odd system sizes.
The scaling of the second order decomposition for a complete time
evolution of $n$ time steps and a total time $T$ with $T = n \cdot dt$
scales as $\mathcal{O}(dt^2)$. Higher orders of the decomposition can
improve the scaling of the error which originates in the non-zero
commutator of the terms in the set $(2k-1,2k)$ and the set $(2k,2k+1)$
at the cost of more terms. Notice that the commutators
$ \left[ \mathcal{L}_{2k-1,2k}, \mathcal{L}_{2k'-1,2k'} \right] = 0$ and
$ \left[ \mathcal{L}_{2k,2k+1}, \mathcal{L}_{2k',2k'+1} \right] = 0$ hold;
thus, the sum of exponentials can be written as a product of exponentials
of two-site terms representing the form used for the efficient numerical
implementation. An alternative to the Trotter decomposition is the
Sornborger decomposition \cite{Sornborger1999}, which has the same
source of error, i.e., splitting a single exponential of non-commuting terms
into multiple exponentials. This decomposition is the one used in the
implementation in \OSMPS{}.

The first and second release of \OSMPS{} use a Krylov subspace
method \cite{Moler2003} for
TEBD, which was argued to be slow \cite{JaschkeMPS} in
comparison to the direct matrix exponential of the Hamiltonian in closed
systems. We now have both methods implemented, with KTEBD and
TEBD taking the matrix exponential of the Hamiltonian or Liouville
operator, respectively. KTEBD follows closely the restriction
of the Krylov method. Moving from Hermitian operators to non-Hermitian
operators, the matrix in the Krylov subspace turns from symmetric tridiagonal
to an upper Hessenberg form. We have to adapt the matrix exponential.
Evidently, for taking the exponential of the Liouville operator, we also have
to choose a matrix exponential for non-Hermitian matrices; we rely on LAPACK's
\texttt{ZGEEV}. The symmetric
MPDO profits from taking the exponentials of the block-diagonal structure.
When building the block-diagonal structure, we ensure that every possible
block is present by adding a $0 \cdot \1$, where $\1$ is the identity
operator of the corresponding subspace. An identity scaled with zeros adds
the information about all present diagonal blocks without altering the
matrix itself.

\subsubsection{Measurements with overlaps                                      \label{sec:otn:mpdo:meas}}

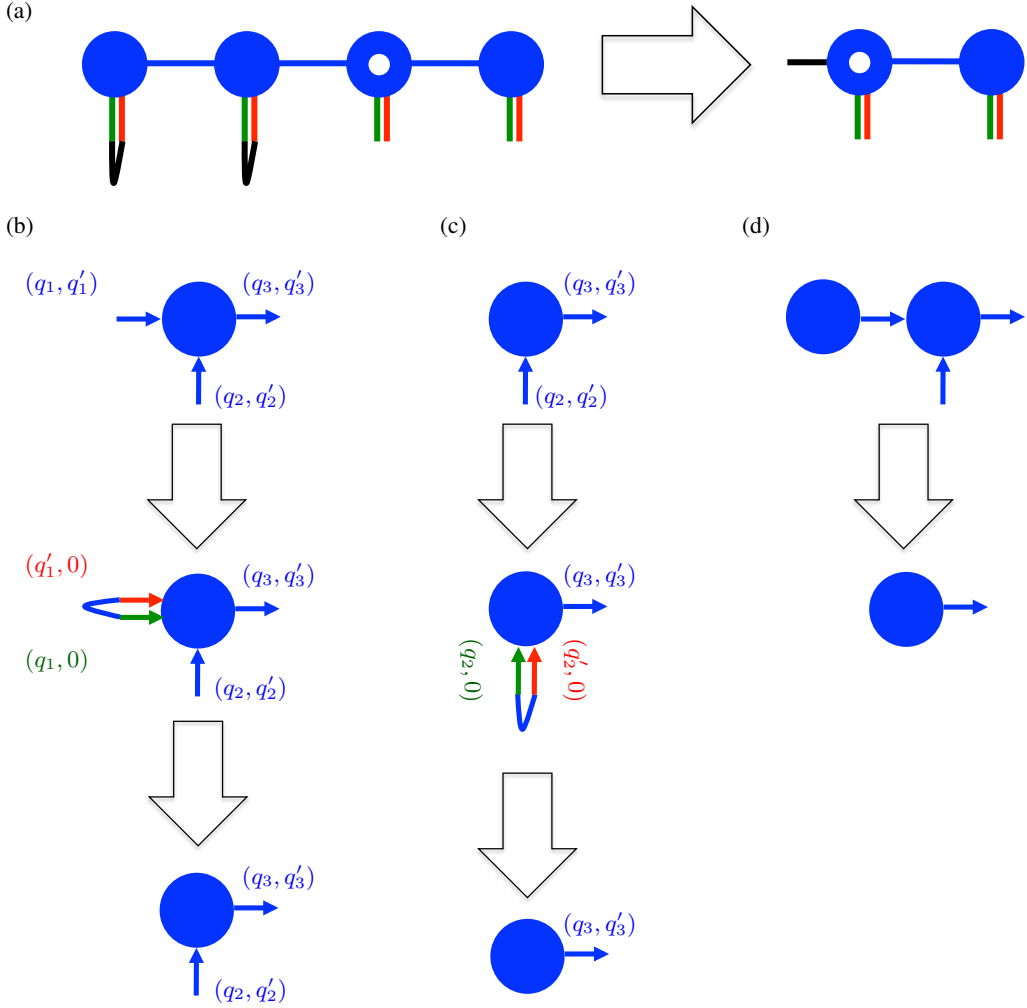
\begin{figure}[t]
 \begin{center}
   \vspace{0.0cm}
   \begin{overpic}[width=0.7 \columnwidth,unit=1mm]{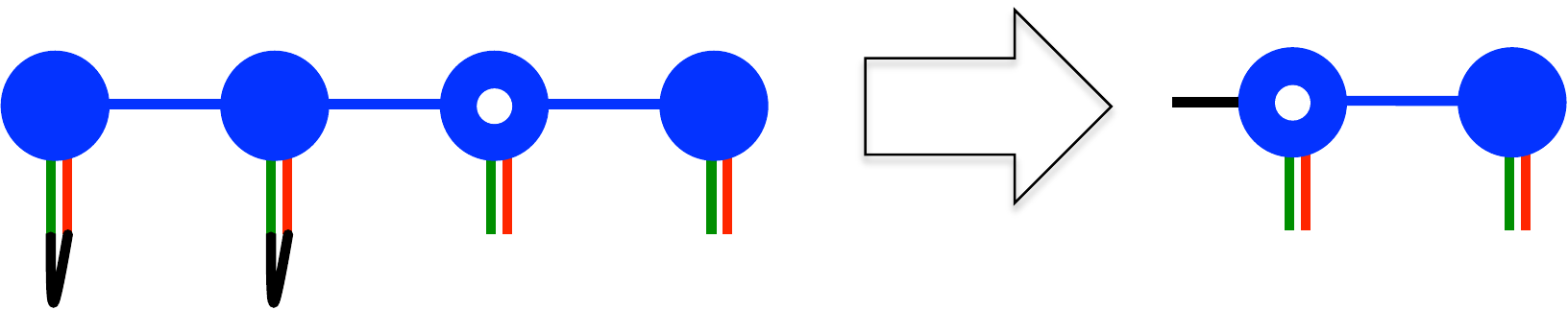}
     \put(-8, 18){(a)}
   \end{overpic}\\
   \vspace{1.2cm}
   \begin{overpic}[width=0.7 \columnwidth,unit=1mm]{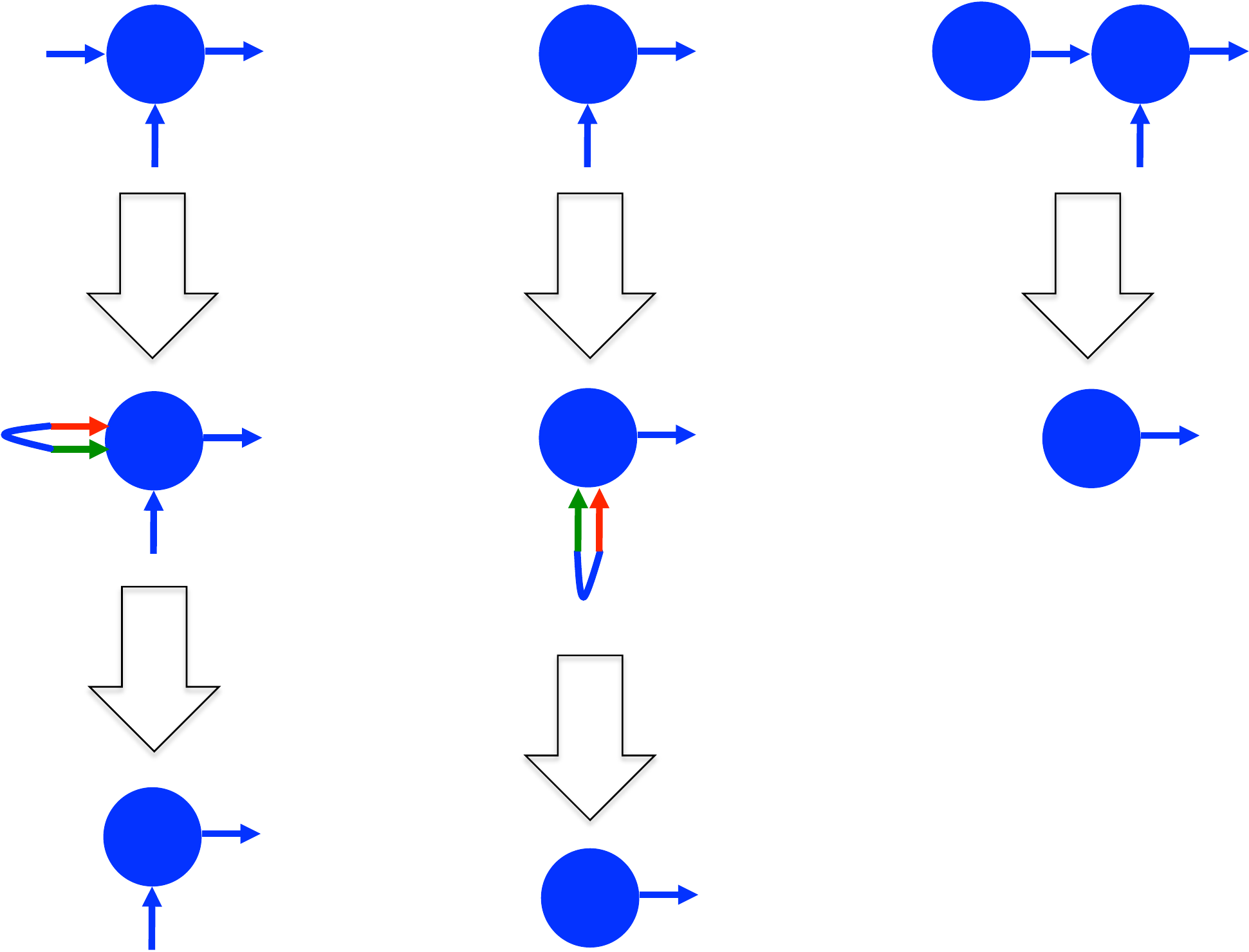}
     \put(-8, 81){(b)}
     \put(38, 81){(c)}
     \put(70, 81){(d)}
     \put(-6, 75){\color{blue}$(q_1, q_1')$}
     \put(14, 63){\color{blue}$(q_2, q_2')$}
     \put(17, 75){\color{blue}$(q_3, q_3')$}
     \put(-6, 45){\color{red}$(q_1', 0)$}
     \put(-6, 35){\color{dg}$(q_1, 0)$}
     \put(14, 32){\color{blue}$(q_2, q_2')$}
     \put(17, 44){\color{blue}$(q_3, q_3')$}
     \put(14, 0){\color{blue}$(q_2, q_2')$}
     \put(17, 12){\color{blue}$(q_3, q_3')$}
     \put(48, 63){\color{blue}$(q_2, q_2')$}
     \put(51, 75){\color{blue}$(q_3, q_3')$}
     \put(40, 38){\rotatebox{270}{\color{dg}$(q_2, 0)$}}
     \put(51, 38){\rotatebox{270}{\color{red}$(q_2', 0)$}}
     \put(51, 44){\color{blue}$(q_3, q_3')$}
     \put(51, 7){\color{blue}$(q_3, q_3')$}
   \end{overpic}
   \caption[Left transfer tensor for an MPDO]
   {\emph{Left transfer tensor for an MPDO.} The left transfer tensor
   for the $k^{\mathrm{th}}$ site is built by taking a partial trace of
   the sites $1, \ldots, k - 1$.
   (a)~We first split and then contract the local Hilbert spaces on the
   sites $1$ and $2$ corresponding to a partial trace, see black curves.
   The results left transfer tensor of rank 1 is also shown in black on
   the right.
   the first site is built with the following steps. (b)~Contract incoming link to
   the left by splitting the link into its original underlying links.
   Then, (c)~Contract the local Hilbert space after splitting it. The
   transfer tensors for sites $k \ge 2$ can be built with subsequent
   steps (c)~Contract the transfer tensor from the site $k - 1$ and
   (d) again.
                                                                                \label{fig:otn:mpdolr}}
 \end{center}
\end{figure}

Measurements of the MPS representation profit immensely from the gauge
installed, which reduces local measurements to operations on a single tensor.
Similarly, correlations are only affected by the sites measured and all sites
in between~\cite{Schollwoeck2011}. Unfortunately, MPDOs do not have such a
benefit, and the complete tensor network has to be contracted to find the
measurement outcome. Therefore, we define a few operations to allow us to obtain those
measures. These operations are similar to the left-right overlaps and transfer
matrices between two states in our MPS algorithms. In detail, we build the
tensors $L_{\alpha}^{[k-1]}$ and $R_{\beta}^{[k+1]}$ such that the local
measurements on site $k$ of the operator $O^{[k]}$ reduce to the contraction
\begin{eqnarray}
  \mathrm{Tr}\left( \rho O^{[k]} \right)
  &=& \sum_{\alpha, i, i', \beta}
  L_{\alpha}^{[k-1]}
  B_{\alpha, (i, i'), \beta}^{[k]}
  O_{(i,i')}^{[k]}
  R_{\beta}^{[k+1]} \, .
\end{eqnarray}
The critical step is the initialization on the left and right end, respectively.
We implement the concept which allows us to use general states, e.g., partially
traced out MPS.\footnote{Such an approach is beneficial for the following example:
(i) Obtain MPS via ground state or time evolution. (ii) Trace out over a subsystem
followed by (iii) a conversion to an MPDO.}
The three major steps are sketched in Fig.~\ref{fig:otn:mpdolr}
for the left side of the tensor network. Figure~\ref{fig:otn:mpdolr} is an
overview of the objective tracing out over site one and two and obtain a rank 1
tensor to be contracted with the rest of the system. First, we split the link to the left
of the MPDO. This step is defined under the assumption that the link was
created as a combined link. The correct decomposition can be stored; this link
cannot change during any MPDO algorithm. If quantum numbers are present as
in Fig.~\ref{fig:otn:mpdolr}(b), we store the conserved quantities in the first
half and fill the second half with zeros. This trick is necessary since
the number of conserved quantities is an attribute of the tensor, not to a
specific link. We contract over the local Hilbert space of the site in the
next step, i.e., Fig.~\ref{fig:otn:mpdolr}(c). The dimensions and quantum
numbers of the original Hilbert space can be obtained from the identity
operator. Transfer tensors for matrices in the bulk of the system are
obtained via a contraction of the transfer tensor on the left, see
Fig.~\ref{fig:otn:mpdolr}(d), followed by the contraction of the local Hilbert
space. We build the right overlap in an analog approach from the other side.
This scheme corresponds in other words to the partial trace of selected sites
in the density matrix.

\subsection{Quantum Trajectories                                               \label{sec:otn:qt}}

Quantum trajectories follow the idea that any mixed quantum state can be
sampled and written via an ensemble of pure states
\begin{eqnarray}
  \rho = \sum_{i} \ket{\phi_i} \bra{\phi_i} \, .
\end{eqnarray}
The essential points of the implementation are the effective, non-Hermitian
Hamiltonian and the sampling over the different trajectories. The latter are
well covered with the data parallelism via MPI present in the package; from
the point where the number of trajectories is specified, they are spread across
all possible cores as any other set of simulations in OSMPS, e.g., ground
state searches for different system sizes. The non-Hermiticity of the
effective Hamiltonian can be addressed analogously to the MPDO cases but on the level
of the Hilbert space and pure quantum states. The effective Hamiltonian
$\Heff$ derived from Eq.~\eqref{eq:otn:lindblad} is
\begin{eqnarray}
  \Heff
  &=& H - \frac{\i \hbar}{2} \sum_{\nu} L_{\nu}^{\dagger} L_{\nu} \, .
\end{eqnarray}
This effective Hamiltonian $\Heff$ violates Hermiticity and
conservation of norm when used in the Schr\"odinger equation. The latter has
its use on the decision on when to apply the Lindblad operators to the system.
The non-Hermiticity leads to changes in the time evolution methods as
discussed in the previous section for the MPDO. The Krylov method, TDVP,
and KTEBD switch from Krylov-Lanczos to Krylov-Arnoldi approximation of
the new state. TEBD uses a function for non-Hermitian matrices
to exponentiate $\Heff$. The LRK method has to use a non-Hermitian
matrix exponential for the local terms but has one additional problem.
LRK itself does not conserve norm which enhances or prevents the loss of
norm due to $\Heff$.

\begin{figure}[t]
 \begin{center}
   \vspace{0.7cm}
   \begin{overpic}[width=0.9 \columnwidth,unit=1mm]{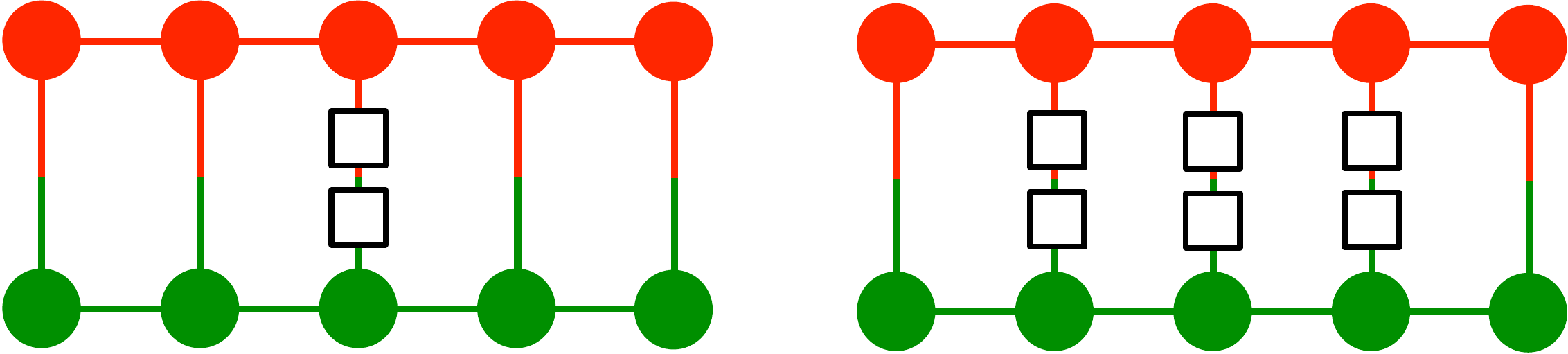}
     \put(-6, 25){(a)}
     \put(50, 25){(b)}
     \put(22.3, 13){$\ast$}
     \put(66.6, 13){$\ast$}
     \put(76.6, 13){$\ast$}
     \put(86.8, 13){$\ast$}
     \put(-6, 19){\color{red} $\bra{\psi}$}
     \put(-6, 2){\color{dg} $\ket{\psi}$}
   \end{overpic}
   \caption[Contraction for unweighted probabilities in quantum
     trajectories]
   {\emph{Contraction for unweighted probabilities in quantum
     trajectories.} (a)~Local Lindblad operators obtain their weight
     by a measurement of norm after contracting it to the current
     quantum state. This step can be reduced to a single site operation.
     (b)~The many-body string Lindblad terms, here shown for three sites,
     need to contract the local operators with the state for each site
     before measuring the norm, i.e., the unweighted probability.
                                                                                \label{fig:otn:qtprob}}
 \end{center}
\end{figure}
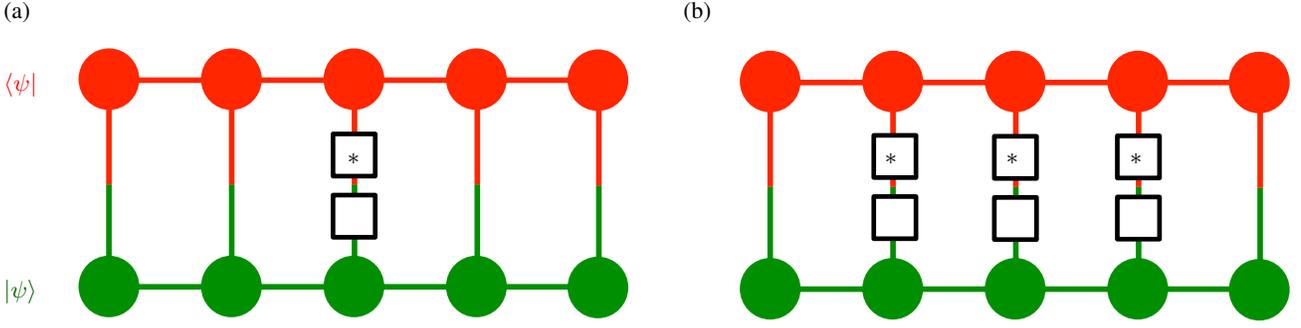

It remains to present the actual algorithm for the trajectories, which is described
in many places~\cite{Bonnes2014}, but reviewed here briefly for completeness. While looping
over the time steps of the evolution, we execute the following steps:
(i)~Draw a random number $\rqt \in \mathcal{U}(0, 1)$ and ensure the state
$\ket{\psi(t=0)}$ is normalized at the beginning. The uniform distribution 
between $0$ and $1$ is written as $\mathcal{U}(0, 1)$. This normalization is necessary as the
decreasing norm of $\ket{\psi(t)}$ is used in the following steps.
(ii)~Calculate time steps under $\Heff$ without renormalizing
the state. (iii)~Measure the norm $\mathcal{N}$ of the state. If
$\mathcal{N} < \rqt$, then we apply a quantum jump according to the steps
(a) through (c) described below. Otherwise, we continue to the next step in the time
evolution, i.e.~(i). The steps to select a random quantum jump are the
following. (a)~Calculate unweighted probabilities for each Lindblad operator
$L_{\nu}$ as $p_{\nu} = \bra{\psi} L_{\nu}^{\dagger} L_{\nu} \ket{\psi}$.
Figure~\ref{fig:otn:qtprob} shows the contractions to obtain the $p_{\nu}$'s
for the two types of Lindblad operators implemented in the package. (b)~We
normalize $p_{\nu}$ to $P_{\nu} = p_{\nu} / \sum_{\nu} p_{\nu}$. (c)~We draw
a random number $r_{\kappa} \in \mathcal{U}(0, 1)$. We apply to the state the
Lindblad operator $L_{\kappa}$ such that
$P_{\kappa - 1} < r_{\kappa} \le P_{\kappa}$ holds. The state
is renormalized after application of the Lindblad operator $L_{\kappa}$ and
we continue the time evolution with step (i).

It is worthwhile to take a look at the
measurement procedure for QTs. Local measurements or correlations can be
averaged over all trajectories with equal weight. For example, a local
observable $\langle O_{k} \rangle$ is calculated for QTs as
\begin{eqnarray}
  \langle O_{k} \rangle &=& \frac{1}{N_{\mathrm{QT}}}
  \sum_{j=1}^{N_{\mathrm{QT}}} \langle O_{k} \rangle_{j} \, ,
\end{eqnarray}
where $N_{\mathrm{QT}}$ is the number of trajectories and
$\langle O_{k} \rangle_{j}$ is the observable of the $j^{\mathrm{th}}$
trajectory. Here, we use the concept that the density matrix is a
statistic ensemble of pure states $\rho = \sum_{j} p_{j} \ket{\psi_j}
\bra{\psi_j}$ with equal probabilities $p_{j} = 1 / N_{\mathrm{QT}}$.
Obviously, this approach does not work for all observables; the purity
$\mathcal{P}$ of each $\ket{\psi_j}$ is equal to one, but the purity of the
density matrix is not necessary the average, i.e., one. We can
calculate the purity as
\begin{eqnarray}
  \mathcal{P} &=&
  \mathrm{Tr} \left[ \rho^2 \right] = \sum_{j,j'=1}^{N_{\mathrm{QT}}}
  \frac{1}{N_{\mathrm{QT}}^{2}} \mathrm{Tr} \left[
  \ket{\psi_{j}} \bra{\psi_{j}} \ket{\psi_{j'}} \bra{\psi_{j'}} \right] \, ,
\end{eqnarray}
where each of the $N_{\mathrm{QT}}^{2}$ traces can be represented as
contraction across a tensor network. But this measurement goes beyond
simple MPI parallelization as the measurement needs all states at the time
of the measurement and a posteriori averaging is not possible. We can
extend this argument to any term with power differing from first order
in $\rho$. Therefore, non-linear measures need all trajectories for
measuring, which can either be achieved with saving each state or while
waiting until all trajectories have reached the measurement.

\subsection{Locally Purified Tensor Networks                                   \label{sec:otn:lptn}}

LPTN time evolution relies on the Trotter decomposition and Kraus operators
applied to the purification of the density, as previously
outlined in Reference~\cite{Werner2016}. Therefore, we restrict the
description to the technical implementations used within \OSMPS{}. We start
with a brief description of the construction and continue with open system
decomposition. We conclude with the setup for finite temperature. The current
implementation is restricted to nearest-neighbor interaction in the
Hamiltonian and local Lindblad operators. Furthermore, no symmetries are
present.

\subsubsection{Construction of Locally Purified Tensor Networks                \label{sec:otn:lptn:states}}

We focus on the conversion of MPS to LPTN and the definition of the infinite
temperature density matrix $\rho_{\infty}$. The conversion from an MPS tensor
$A_{\alpha, i, \beta}^{[k]}$ representing site $k$ to an LPTN tensor is as
simple as inserting an additional link of dimension $1$, i.e.,
$B_{\alpha, i, \kappa, \beta}^{[k]}$. The local tensors for the definition
of $\rho_{\infty}$ have the link dimension $(1, d, d, 1)$ and are diagonal
in the entries
\begin{eqnarray}
  B_{\alpha, i, \kappa, \beta}^{[k]} &=& \delta_{i, \kappa} \, .
\end{eqnarray}

\subsubsection{Trotter decomposition for open systems                          \label{sec:otn:lptn:states}}

The evolution of an LTPN separates the Hamiltonian part of the Liouville
operator $\mathcal{L}$ from the dissipative part $\mathcal{D} = \sum_{\nu}
L_{\nu} \rho L_{\nu}^{\dagger} - \frac{1}{2} \{L_{\nu}^{\dagger} L_{\nu},
\rho \}$ in a second order Trotter decomposition. The exponential of the
time evolution is approximated with
\begin{eqnarray}
  \mathrm{e}^{\mathcal{L} dt}
  \approx \mathrm{e}^{- \frac{\i}{\hbar} [H, \rho] \frac{dt}{2}}
          \mathrm{e}^{\mathcal{D} dt}
          \mathrm{e}^{- \frac{\i}{\hbar} [H, \rho] \frac{dt}{2}} \, .
\end{eqnarray}
The Hamiltonian contribution evolves each part of the purification
$\rho = X X^{\dagger}$, where it is sufficient to propagate $X$. The
Hamiltonian itself is approximated with two-site propagators in a Sornborger
decomposition~\cite{Sornborger1999}. We remain with the dissipative part,
which we express as Kraus operators:
\begin{eqnarray}
  \mathrm{e}^{\mathcal{D} dt} \Lket{\rho}
  &=& \sum_{\nu'} K_{\nu'} X X^{\dagger} K_{\nu'}^{\dagger}
   =  \sum_{\nu'} (K_{\nu'} X) (K_{\nu'} X)^{\dagger} \, . \nonumber \\
\end{eqnarray}
We recall that the local Kraus operators keep the locally purified tensor network
in its form. We generate the Kraus operators from the first order
approximation~\cite{Cappellaro2012}
\begin{eqnarray}
  \mathrm{e}^{\mathcal{D} dt} \Lket{\rho}
  &=& \sum_{\nu' = 0}^{N} K_{\nu'} \rho K_{\nu'}^{\dagger} + \mathcal{O}(dt) \, , \\
  K_{\nu' = 0} &=& \1 - \frac{dt}{2} \sum_{\nu=1}^{N} L_{\nu}^{\dagger} L_{\nu} \, , \\
  K_{\nu} &=& \sqrt{dt} L_{\nu} \, , \nu = 1, \ldots, N \, .
\end{eqnarray}
Alternatively, we can decompose the exponential of the Liouville operator
into Kraus operators. The contraction of the Kraus operators increases the
auxiliary dimension $\kappa$ to a larger $\kappa' = \kappa (N + 1)$. The
dimension is reduced in a truncation similar to the truncation when splitting
nearest neighbor sites.

\subsubsection{Imaginary time evolution for finite-T states                    \label{sec:otn:lptn:imagt}}

The finite temperature states according to the Gibbs distribution are
generated in an imaginary time evolution starting from the infinite
temperature state $\rho_{\infty}$ as explained in Sec.~\ref{sec:otn:mpdo:imagt}.
As evolution is purely Hamiltonian and does not contain any dissipative part
$\mathcal{D}$, we can reuse the decompositions for MPS/MPDO methods. Both
second and fourth order Sornborger approximations are available.

\section{Simulation Setup and Convergence                                      \label{sec:otn:conv}}

We describe the convergence of the different scenarios in the next three
subsections. We start with the analysis of the finite-T states in
Sec.~\ref{sec:otn:conv:finitet}; we use the quantum Ising model as example.
Then, we move to the Lindblad master equation and consider simulations
without and with symmetry. Section~\ref{sec:otn:conv:oqsa} considers the
dissipative dynamics of an exciton for the convergence study of a system
without conserved symmetry. The transient dynamics of the Bose-Hubbard model
with number conservation are described in Sec.~\ref{sec:otn:conv:oqsb}.

\subsection{Finite-T states in the Ising model (MPDO and LPTN)                 \label{sec:otn:conv:finitet}}

We now turn to finite temperature states according to the Gibbs distribution
and their convergence. We compare LPTNs and MPDOs; quantum trajectories cannot
simulate finite temperature states. We take the quantum Ising model defined
in Eq.~\eqref{eq:otn:HQI} as an example. The comparison is restricted
to the TEBD2 time evolution methods and the overall Hilbert space
without addressing a specific sector in the $\mathbb{Z}_2$ symmetry.

\begin{figure}[t]
 \begin{center}
   \vspace{0.7cm}
   \begin{minipage}{0.47\linewidth}
      \vspace{0.05cm}
      \begin{overpic}[width=1.0 \columnwidth,unit=1mm]{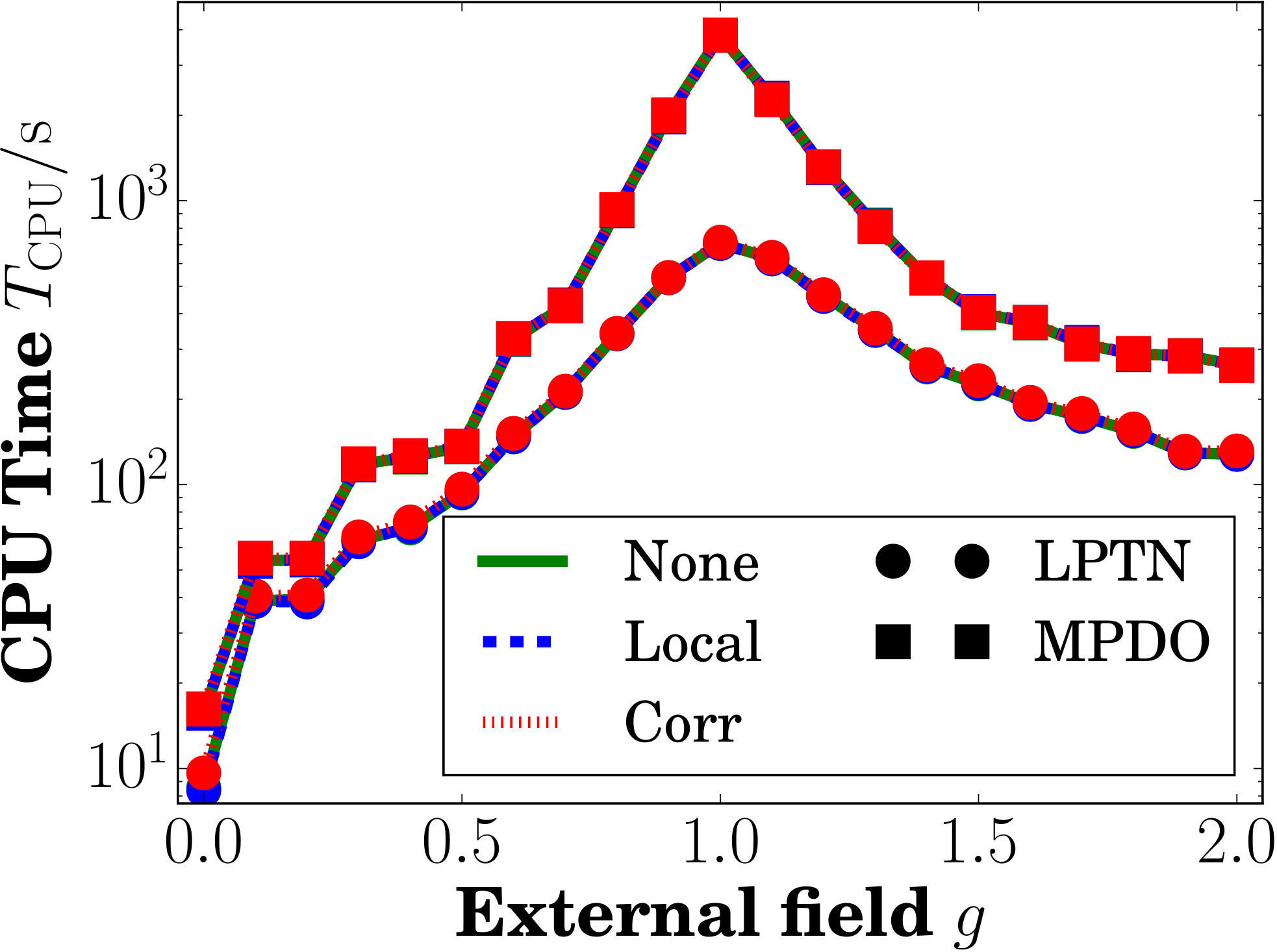}
        \put(2, 79){(a)}
      \end{overpic}
    \end{minipage}\hfill
    \begin{minipage}{0.47\linewidth}
      \vspace{0.05cm}
      \begin{overpic}[width=1.0 \columnwidth,unit=1mm]{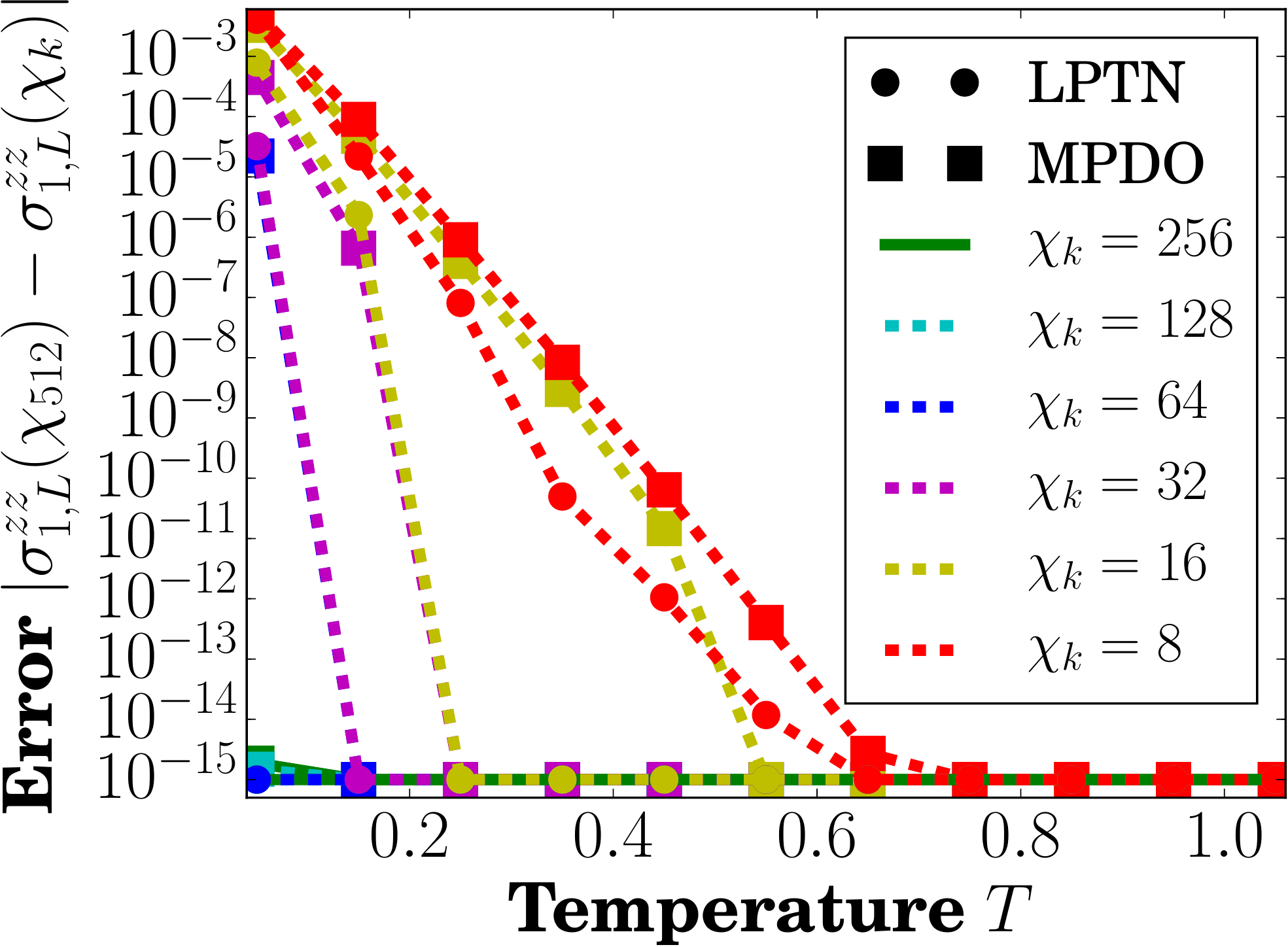}
        \put(2, 79){(b)}
      \end{overpic}
    \end{minipage}
   \caption[Performance of finite-T simulations]
   {\emph{Performance of finite-T simulations.} We consider the
     quantum Ising model without $\mathbb{Z}_{2}$ symmetry to study
     basic convergence of the finite-T states represented by LPTNs and
     MPDOs. The system size for the simulations is $L = 100$.
     (a)~The CPU times $\Tcpu$ for different measuremts indicate that
     LPTNs are favorable for this model. The bond dimension is
     $\chi = 512$. The measurements are not the major cost and all three
     curves for ``None", ``Local", and ``Corr" overlap.
     (b)~We compare the error of the $zz$-correlation measurement between the
     first and last site to the most accurate result with $\chi = 512$.
     Data points, especially circles for MPDOs, may hide behind other
     data points of lower bond dimension for the error level $10^{-15}$.
     We conclude from the lower errors for the same bond dimension that
     LPTNs use their bond dimension more efficiently in this example.
                                                                                \label{fig:otn:finiteT_study}}
 \end{center}
\end{figure}

Figure~\ref{fig:otn:finiteT_study} answers two important questions: which method
needs less CPU time to run the simulations and which network type uses its
bond dimension more efficiently? The first answer is important to use
computational resources in the best possible way whenever both tensor
networks can be applied to the problem. The second question can deliver a
partial answer as to which is the most favorable network to store entanglement for any
state similar to those finite-T states. We consider the finite-T evolution
with the first measurements at $T = 1.05$ and the last measurement at
$T = 0.05$ in steps of $\Delta T = 0.1$. We use a time step $dt = 0.01$,
and additional steps $dt' < dt$ to reach the measurements at the corresponding
temperatures. The temperatures are in units of the interaction $J$, and we have
a unitless $k_B = 1$.

We first discuss
Fig.~\ref{fig:otn:finiteT_study}(a) and the computational effort for the imaginary
evolution and the measurements. We choose a system size of $L = 100$, a
maximal bond dimension $\chi = 512$, and three type of measurements:
(i) only default measurements as the energy; and (ii) default measurements and two
local observables, i.e., $\sigma_{j}^{x}$ and $\sigma_{j}^{z}$. (iii) The last
type of simulations computes the default measurements, the local measurements
presented previously, and the correlations of the Pauli matrices in the $x$- and
$z$-direction. All simulations are carried out on a \miosixxeightx{}.
The peak of the CPU time $\Tcpu$ is located around the critical point; the
entanglement is maximal at the critical point of the Ising model, i.e., 
$g = 1.0$.
We see a clear difference between the LPTN and MPDO times for no measurements,
which is related in aspects to the computational scaling of the underlying
operations. LPTNs build the two-site Hamiltonian on a $d^2 \times d^2$ matrix,
where the local dimension $d$ is equal to two in the qubit example. In contrast,
the Hamiltonian represented in Liouville space is of dimension $d^4 \times d^4$.
The contraction of a single two-site propagator with a two-site tensor then
scales as $\chi^2 d^6$ for the LPTN and $\chi^2 d^8$ for the MPDO. We remark
that our MPDO implementation could be further optimized using the fact
that the commutator $\left[ H \otimes \1, \1 \otimes H^{T} \right] = 0$. Thus,
we can simplify the action of the two site propagator in Liouville space to
$\exp\left(H \otimes \1 + \1 \otimes H^{T}\right) = \left(\exp(H) \otimes \1\right)
\left(\1 \otimes \exp\left(H^{T}\right)\right)$. Each exponential can be applied separately
in two steps with a scaling of $\chi^2 d^6$ for each contraction. However, this approach
would in our eyes defeat the purpose of an MPDO, which is representing the
density matrix as an MPS; but here, the link of the local dimension has to
remain split throughout the complete evolution. Synergy effects between the MPDO
and MPS implementation would be decreasing.
The measurements, executed in addition to the imaginary time evolution, increase
computation time where they are beyond fluctuations. LPTNs tend to have less
cost for measurements as local and quasi-local measurements do not have to
contract the complete tensor network. But overall, the selected measurements
are not the significant costs in these simulations.

Figure~\ref{fig:otn:finiteT_study}(b) describes the error for different bond dimensions
in comparison to the most accurate bond dimension. We choose this most accurate
bond dimension as $\chi_{\max} = 512$. The system size is $L = 100$, and
we set the transverse field to $g= 1.0$ maximizing the possible entanglement in the
quantum Ising model. The correlation is among the observables most affected by a
limited bond dimension; therefore we consider
\begin{eqnarray}
  \ezz &=& | \langle \sigma_{1}^{z} \sigma_{L}^{z} \rangle_{\chi_{\max}}
           - \langle \sigma_{1}^{z} \sigma_{L}^{z} \rangle_{\chi} | \, ,
\end{eqnarray}
where $\chi < \chi_{\max}$. We see a decreasing error $\ezz$ as the bond
dimension $\chi$ approaches $\chi_{\max}$ for both LPTN and MPDO. This
behavior is expected. But we also observe that each curve
representing the LPTN has a smaller error than the corresponding MPDO curve.
Assuming that the simulations with the highest bond dimension are both
converged and yield the correct results, LPTNs use the bond dimension more
efficiently than MPDOs. This statement is supported by the error in the
energies for the the final temperature. The LPTN has a maximal absolute error
of $2.09 \cdot 10^{-5}$ for $L = 100$, $\chi = 512$, $g = 1.0$, while the error for
the MPDO yields $6.21 \cdot 10^{-4}$. Thermal energies for low enough
temperatures can be calculated via the eigenvalues of Majorana Hamiltonian
after the Jordan-Wigner transformation; energies of excited states $E'$ are
truncated if $\exp(- (E' - E_{0}) / T) < 10^{-16}$ and avoids iterating over
all $2^L$ eigenenergies.
This effect might also contribute to the CPU times
observed. Lower bond dimension leads to the use of less computational resources.
Unlike a variational or imaginary time ground state search, we cannot
stop the algorithm at a defined precision and compare the CPU times.

One may ask where the speed-up of the LPTN in comparison to the MPDO
occurs and how it scales with system size. This question is in relation
with the known growth of entanglement associated with quantum critical
phenomena \cite{Osterloh2002,Vidal2003b}. We consider the system sizes
$L \in \{ 100, 150, 200 \}$ at the critical point $g = 1.0$ for the scenario
with local measurements at bond dimension $\chi = 512$. We compare the
the ratio $r$ of MPDO over LTPN from measurement 1 to 10 and 10 to 11 according
to the time stamp of the results files. We obtain the ratios $r(L) = \{
4.17, 4.38, 4.36 \}$ for ascending system sizes, and $r(L) =  \{5.52, 6.00, 6.13\}$
are the ratios between the measurements 10 and 11. We observe the trend that
the speed-up has a larger contribution at the end of the time-evolution
and increases with system size. We point out that the major part of the
time evolution takes place for the last data points due to the inverse
relation of temperature to evolution time.
%
%
In conclusion for the finite-T simulations, LPTNs scale better in the
scenario of the quantum Ising model. The maximal difference in CPU time
is around the critical point and suggests that LPTNs support
entanglement better for imaginary time evolutions.

\subsection{Local Lindblad operators without symmetry (QT, MPDO, and LTPN)     \label{sec:otn:conv:oqsa}}

\begin{figure}[t]
 \begin{center}
   \vspace{0.7cm}
   \begin{minipage}{0.47\linewidth}
      \vspace{0.05cm}
      \begin{overpic}[width=1.0 \columnwidth,unit=1mm]{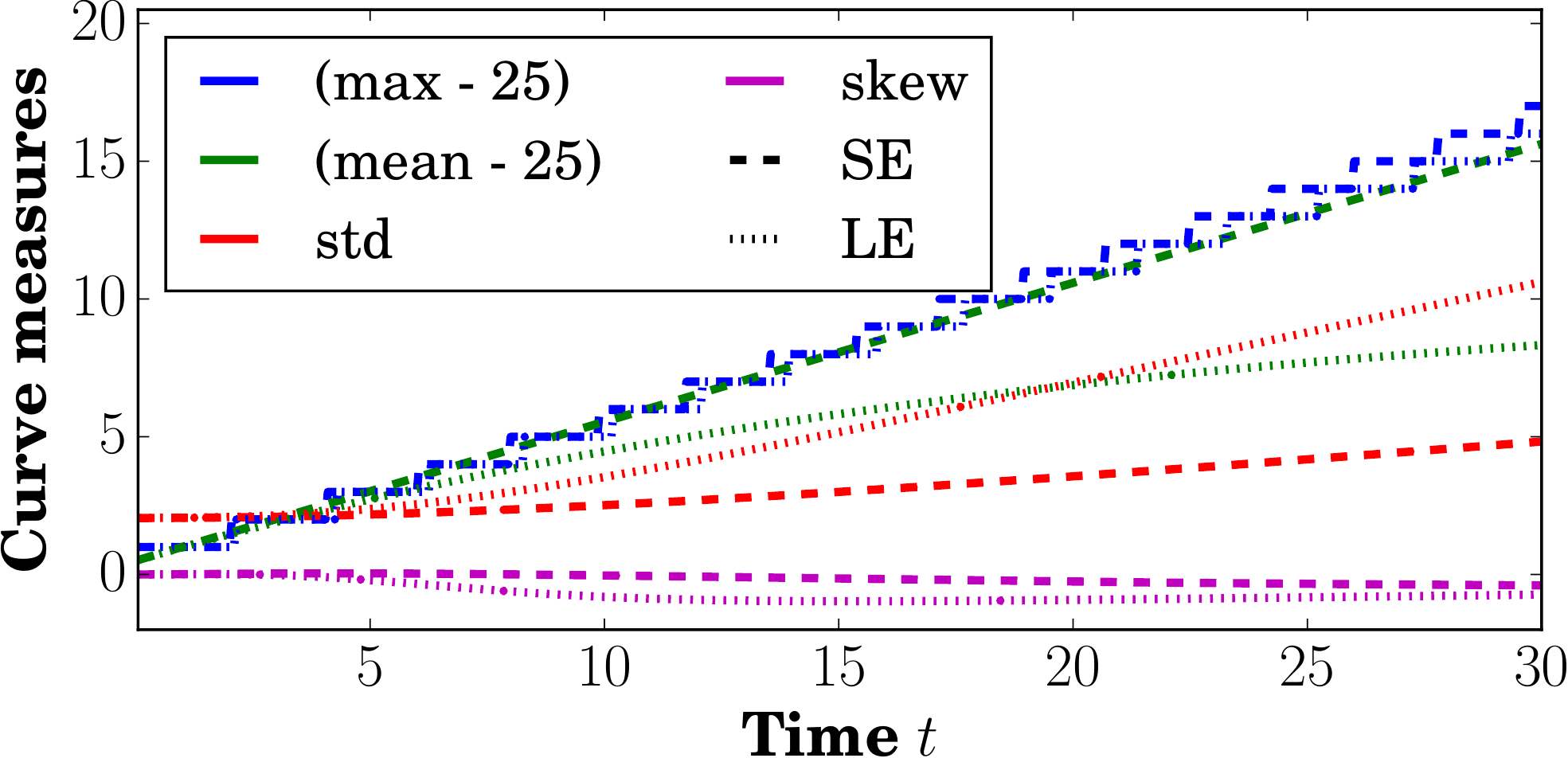}
        \put(2, 54){(a)}
      \end{overpic}
    \end{minipage}\hfill
    \begin{minipage}{0.47\linewidth}
      \vspace{0.05cm}
      \begin{overpic}[width=1.0 \columnwidth,unit=1mm]{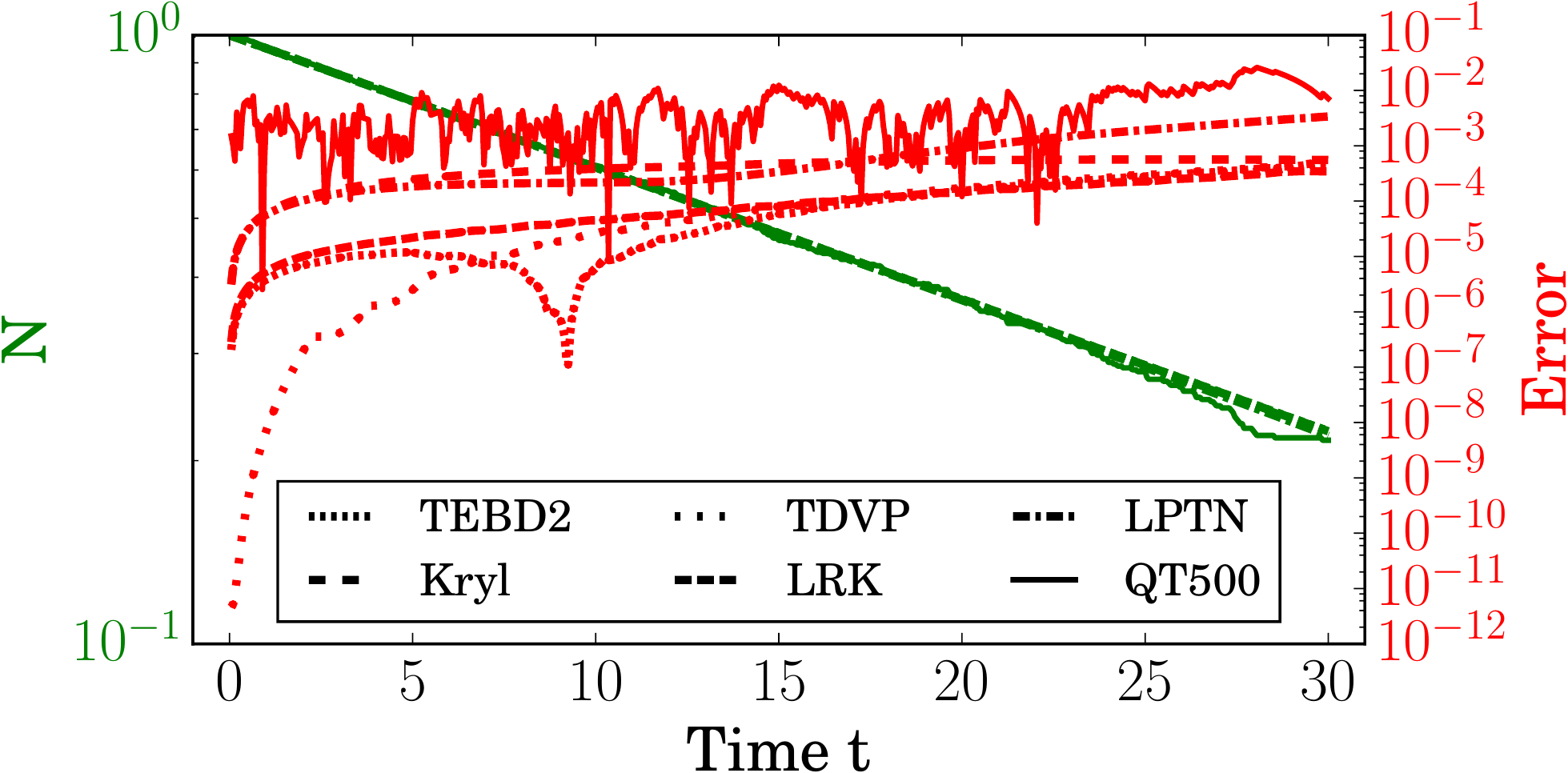}
        \put(2, 54){(b)}
      \end{overpic}
    \end{minipage}\vspace{0.5cm}
   \begin{minipage}{0.47\linewidth}
      \vspace{0.05cm}
      \begin{overpic}[width=1.0 \columnwidth,unit=1mm]{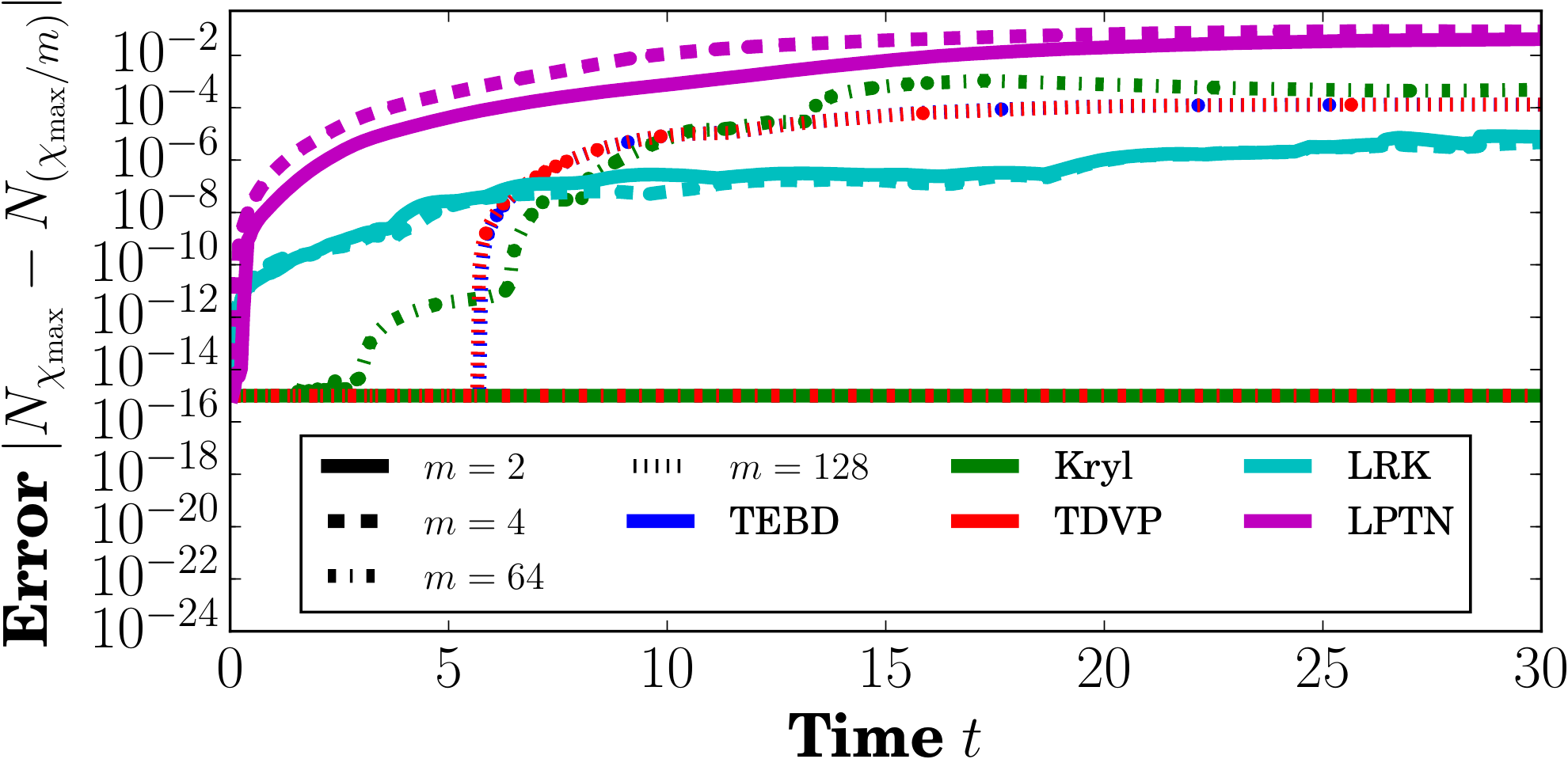}
        \put(2, 54){(c)}
      \end{overpic}
    \end{minipage}\hfill
    \begin{minipage}{0.47\linewidth}
      \vspace{0.05cm}
      \begin{overpic}[width=1.0 \columnwidth,unit=1mm]{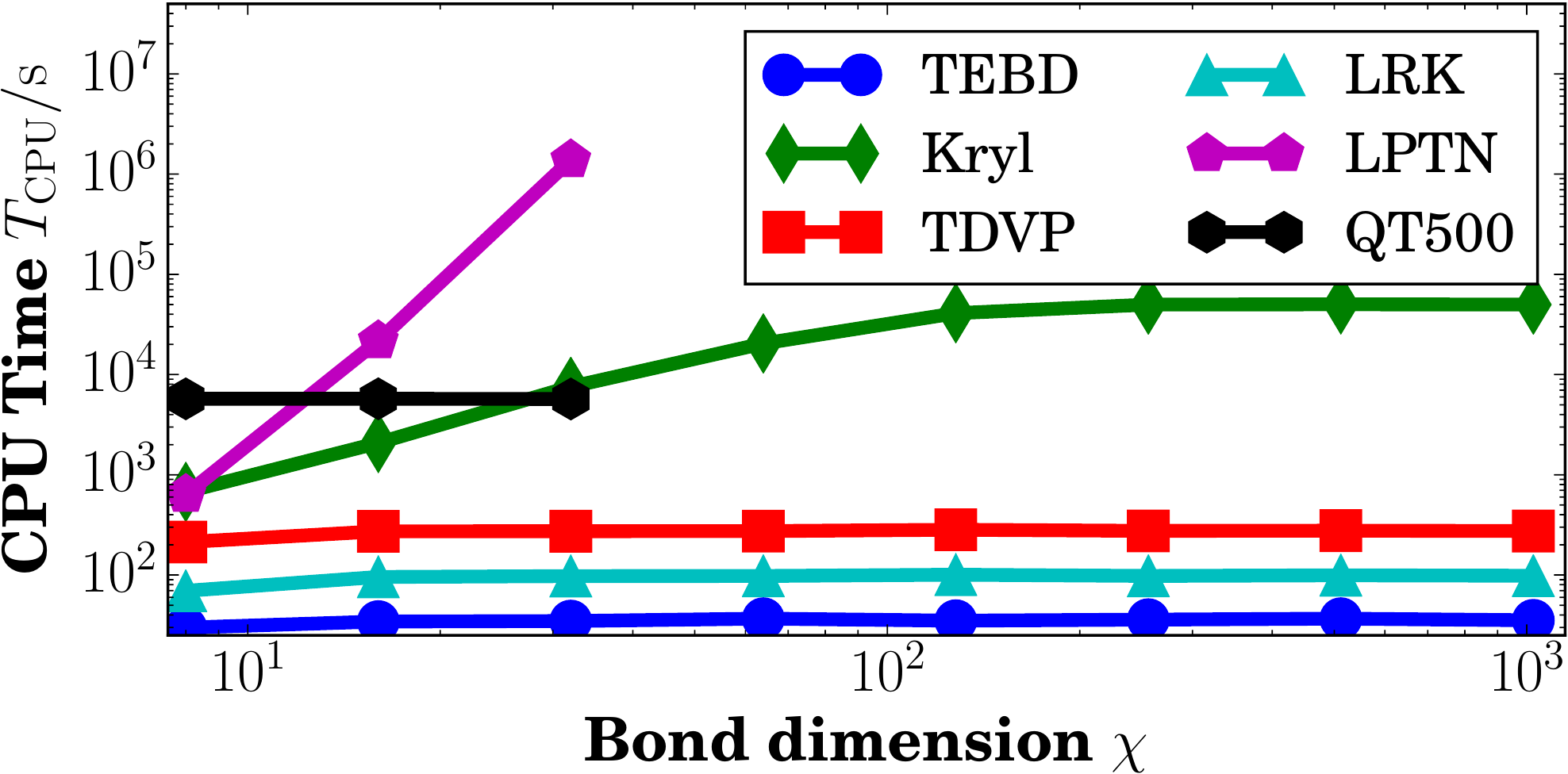}
        \put(2, 54){(d)}
      \end{overpic}
    \end{minipage}
   \caption[Convergence of open quantum system exciton dynamics]
   {\emph{Convergence of open quantum system exciton dynamics.}
     (a)~The different momenta of the exciton capture best the effects of
     the dephasing. Although the maximum (max) of the exciton almost stays
     constant, the mean (mean), standard deviation (std), and skewness (skew)
     capture the dynamics of the Lindblad equation (LE) with coupling
     $\lambda = \lambda_{d} = 0.05$ very well, shown side-by-side with the
     Schr\"odinger equation (SE).
     (b)~The coupling to the decay Lindblad operator induces an exponential
     decay of the total number of particles in the excited state. All time
     evolution methods reproduce this exponential decay (green); thus, we
     show the corresponding error (red). The TDVP for MPDO works best.
     (Linestyles in the legend are used for green and red curves.)
     (c)~We show how efficiently the bond dimension is used by comparing
     the convergence of $N(t)$ for the different bond dimension to the
     maximal bond dimension, i.e., $\chi_{\max}^{\mathrm{MPDO}} = 1024$
     and $\chi_{\max}^{\mathrm{LPTN}} = \kappa_{\max} = 32$. MPDO
     simulations with TEBD and TDVP use the bond dimension very efficiently
     for this problem and already show no error for $\chi = 16$ in comparison
     to $\chi_{\max}$.
     (d)~From a pure resource perspective, TEBD is favorable over LRK
     over TDVP and over Krylov when using MPDOs. LPTN-TEBD also
     scales unfavorable, but the additional link for each site with
     $\chi = \kappa$ induces matrices with dimension $\chi^2$ rather
     than $\chi$ affecting the scaling of operations.
                                                                                \label{fig:otn:excitonconv}}
 \end{center}
\end{figure}

Lindblad equations without conserved quantities and local channels have a
variety of applications reaching from the quantum Ising model with local
spin flips over XXZ model transport problems using local channels at both
end of the chains \cite{Prosen2011} to lossy photon cavities. We choose to simulate the
transport of an exciton with the initial condition and Hamiltonian based
on reference \cite{Lusk2015,Zang2017}. Examples of the Lindblad operators for
transport problems can be motivated from \cite{Plenio2008}, which describes
energy loss and dephasing noise in molecular structures. We consider
the Hamiltonian
\begin{eqnarray}
  H &=& J \sum_{k=1}^{L-1} b_{k} b_{k+1}^{\dagger} + h.c.
        + \Delta \sum_{k=1}^{L} n_{k}
        - \mu \left(b_{k}^{\dagger} + b_{k} \right) \, ,
\end{eqnarray}
and $J$ is the tunneling strength, $\Delta$ an on-site potential, and
$\mu$ the driving due to the electromagnetic field. The latter is turned
off on all sites for all simulations, i.e., $\mu = 0$. The corresponding
Lindblad master equation with loss of a strength $\gamma$ and
dephasing, coupled with $\gamma_{d}$, is defined as
\begin{figure}[t]
 \begin{center}
   \vspace{0.7cm}
   \begin{minipage}{0.47\linewidth}
      \vspace{0.05cm}
      \begin{overpic}[width=1.0 \columnwidth,unit=1mm]{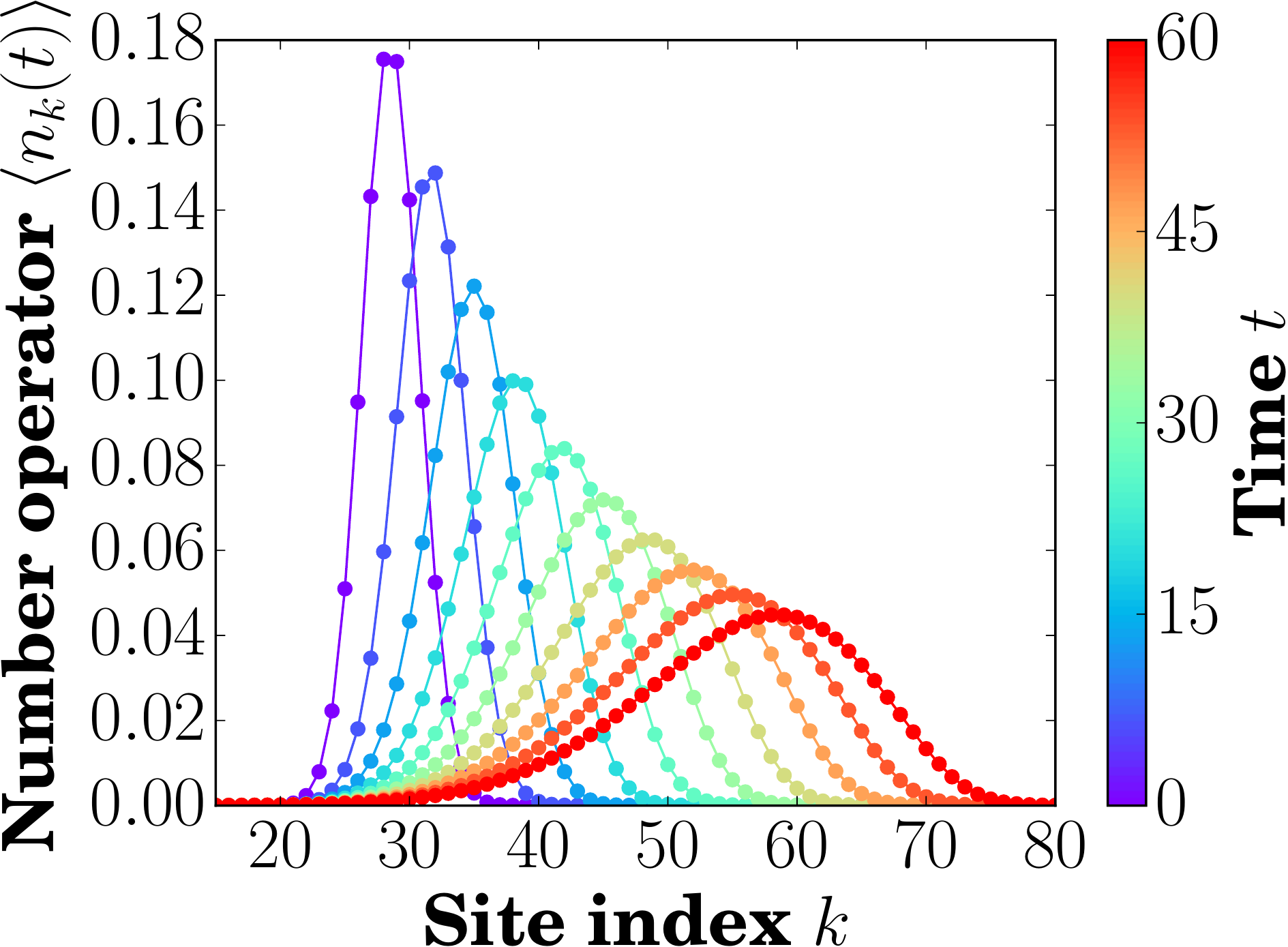}
        \put(2, 84){(a)}
      \end{overpic}
    \end{minipage}\hfill
    \begin{minipage}{0.47\linewidth}
      \vspace{0.05cm}
      \begin{overpic}[width=1.0 \columnwidth,unit=1mm]{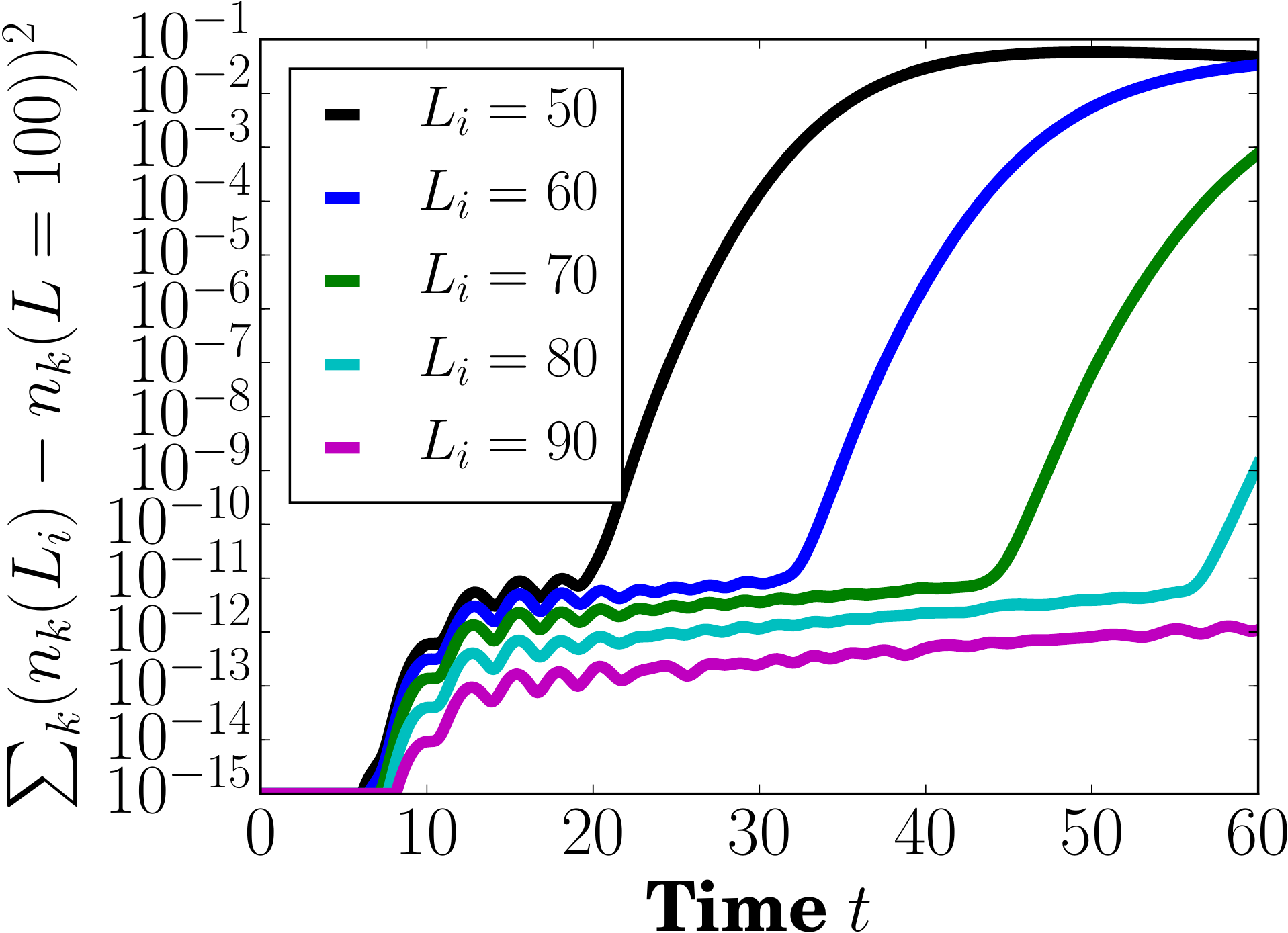}
        \put(2, 82){(b)}
      \end{overpic}
    \end{minipage}
   \caption[Exciton dynamics with open boundary conditions]
   {\emph{Exciton dynamics with open boundary conditions.}
     (a)~The dynamics of the largest system with $L = 100$. The initial wave
     package loses height while spreading out.
     (b)~We consider different system sizes $L$ to estimate when the reflection
     of the exciton returns from the boundary at site $k = L$. The
     exciton is initially in the sites $k = 11, \ldots, 40$ and moving right.
     System sizes of $L = 60$ ($L = 90$) are sufficient to simulate the closed
     system for a time of $t = 30$ ($t = 60$). Times are in the unit of the local
     term $\Delta$.
                                                                                \label{fig:otn:exciton}}
 \end{center}
\end{figure}
\begin{eqnarray}
  \frac{\partial}{\partial t} \rho
  &=& - \frac{\i}{\hbar} \left[ H, \rho \right]
      + \sum_{k=1} \gamma \left( b_{k} \rho b_{k}^{\dagger}
      - \frac{1}{2} \left\{ b_{k}^{\dagger} b_{k}, \rho \right\} \right)
      + \sum_{k=1} \gamma_{d} \left( n_{k} \rho n_{k}
      - \frac{1}{2} \left\{ n_{k} n_{k}, \rho \right\} \right) \, .
\end{eqnarray}
We define the initial state $\ket{\psi(t=0)}$ as a product state on three
subsystems $A$, $B$, and $C$, i.e., $\ket{\psi(t=0)} = \ket{\psi(t=0)}_{A}
\otimes \ket{\psi(t=0)}_{B} \otimes \ket{\psi(t=0)}_{C}$. The subsystems
$A$ and $C$ are in the vacuum state $\ket{\psi(t=0)}_{A} = \ket{0 \cdots 0}$
and $\ket{\psi(t=0)}_{C} = \ket{0 \cdots 0}$, respectively. Subsystem $A$
contains $L_A$ sites; subsystem $C$ spans $L_C$ sites. The initial exciton
is defined via the relation from \cite{Zang2017} on the subsystem $B$ with
a size of $L_B$ sites:
\begin{eqnarray}                                                               \label{eq:otn:exct0}
  \ket{\psi(t=0)}_{B} &\propto&
  \sum_{k=L_A + 1}^{L_A+L_B} \mathrm{e}^{\i k_0 k a}
  \mathrm{e}^{-a^2(k - k_0)^2 / (2 \sigma^2)} b_{k}^{\dagger}
  \ket{0 \cdots 0} \, . \nonumber \\
\end{eqnarray}
Thus, the total system has a system size of $L = L_A + L_B + L_C$ sites. The
exciton on subsystem $B$ is not a product state and has entanglement. It
can be constructed on the completed Hilbert space and decomposed
via SVDs to an MPS. This approach is limited to about $24$ sites and supported
by the OSMPS Python interface. On the other hand, we can write the state
as a sum of MPSs, where each state in the sum of Eq.~\eqref{eq:otn:exct0} has
a bond dimension of 1. We compress this MPS yielded by the summation, which
scales better with larger $L_{B}$.
Figure~\ref{fig:otn:exciton}(a) pictures the exciton at different points in
time and we observe a loss of height while spreading out in the closed
system. 
We are well aware that this setup of the problem has problems
from the back reflection of the exciton at the boundary $k = L$. Therefore,
Fig.~\ref{fig:otn:exciton}(b) shows the deviation of the exciton's mean
position for different system sizes for the closed system over time in
comparison to $L = 100$. We see small errors at the order of $10^{-12}$
and do not concentrate on these errors as $10^{-12}$ is at the order
of truncated singular values. We conclude that a system size of $L = 90$
is sufficient for an evolution
time $\tau = 60$ without considering reflections from the boundary.
To estimate the error, we look at the number operator
$N(t) = sum_{k=1}^{L} n_{k}(t)$ as
\begin{eqnarray}                                                                \label{eq:otn:expexciton}
  \epsilon_{N(t)} = \left| N(t) - N(0) \exp(- \gamma t) \right| \, ,
\end{eqnarray}
where the relation $dN(t) / dt = - \gamma N(t)$ can be easily derived when
considering the commutation relation between $b_{k}$, $b_{k'}^{\dagger}$ and
$n_{k''}$ as well as $d / dt N(t) = d / dt \mathrm{Tr} \left[ N \rho(t) \right]
= d / dt \mathrm{Tr} \left[ N d / dt \rho(t) \right]$. This
property is one result which can be checked besides convergence regarding
the bond dimension or other convergence parameters. We start with an analysis
of the momentum of the exciton; the position of the maximum of the exciton,
the mean, the standard deviation, and the skewness give a good impression of
the dynamics, see Fig.~\ref{fig:otn:excitonconv}(a). We observe that for a coupling
$\lambda = \lambda_{d} = 0.05$ the position of the maximum remains almost equal
to the closed system case. The mean, i.e., "center of mass", propagates at
a slower speed in the open system in comparison to the closed system. This trend
is reflected in the increased standard deviation and skewness of the exciton.
These measures can be used for a more detailed analysis of the open system
dynamics of the exciton, but we turn to the evaluation of the different tensor
networks and algorithms.

Figure~\ref{fig:otn:excitonconv}(b) shows the error in
the number of excitations, which decays exponentially according to
Eq.~\eqref{eq:otn:expexciton}. TDVP has the lowest error; TEBD has a local minimum
of the error around $t \approx 10$ switching between under- and overestimating
the number of excitations in the system and recovers the envelope otherwise in
our understanding. The LRK has an error similar to TEBD. In the long time limit,
all of these three algorithms using MPDOs have a similar error. The Krylov method
has an error about an order of magnitude bigger at the beginning and recovers the
error of the other algorithms in the long-time limit; it is not considered an
option in our opinion. If we consider the second
tensor network, LTPN, the error is the largest in the long-time limit. If we
consider in addition the CPU times in Fig.~\ref{fig:otn:excitonconv}(d) for each
simulation as a function of the maximal bond dimension, we can get a much better
picture as to what method is best suited to the problem. Two things become evident. (i) The Krylov
method for MPDOs is two orders of magnitude slower than the next slowest algorithm
for MPDOs. The other algorithms reach a saturation of the CPU time, and we conclude
that they do not exceed the maximal bond dimension. Otherwise, TEBD is preferable
when applicable. For long-range interactions in the Hamiltonian, LRK is preferable
over TDVP from a resource viewpoint. (ii) The LPTN algorithm requires much higher
resources in comparison to MPDO methods. We remind the reader that a maximal bond
dimension $\chi_{\max} = \kappa = 32$ leads to a bond dimension of $32^2$ in many
tensor operations due to the auxiliary link in the local tensors of the same
dimension. This approach is fair when comparing the splitting of a two-site tensor
when $\chi_{\max,\mathrm{MPS}} = \chi_{\max,\mathrm{MPDO}} \kappa_{\max}$; in both
cases, we split a matrix of $\chi_{\max,\mathrm{MPS}} d \times
\chi_{\max,\mathrm{MPS}} d$. Thus, we can compare the LPTN data for $\chi = 16$ with
the MPDOs of $\chi = 256$. The error does not outweigh the higher CPU time. It
remains the question if novel methods for LPTNs can overcome the problem
\cite{Wegener2018}, which also consider the optimal ratio between $\chi$ and
$\kappa$ for LPTNs.
%

Finally, Figure~\ref{fig:otn:excitonconv}(c) answers in parts how efficiently the
different algorithms use their bond dimension. We compare the values of the
number operator for MPDOs with $\chi_{\max} = 1024$ and LTPNs with
$\chi_{\max} = \kappa = 32$ to simulation with lower bond dimension and get
a more detailed picture of convergence. MPDOs profit from using a very low
bond dimension for TDVP and TEBD. We remark that this model including this
initial condition has very low requirements with regards to the bond
dimension with MPS and MPDO methods. In fact, simulations with
$\chi_{\max} / 64 = 16$ are already not distinguishable from simulations
with $\chi_{\max}$. In contrast, the LRK method shows, despite the same
error in the total number of excitations, an error between bond dimensions
$512$ and $1024$, i.e., they must exceed bond dimensions up to and beyond
$512$. The Krylov method does not use the full bond dimension according to
this data, but is less efficient in using the bond dimension in comparison
to TEBD and TDVP. The LTPN algorithm has the highest error for this
comparison, although the bond dimension for both links is with $32$ much
lower. Saturating both links leads to the larger CPU times and the higher
error if values are truncated throughout the simulation. The inefficient
use of the bond dimension in the LPTN is related to the loopy network
structure when looking at a the complete network representing $\rho$ instead
of its purification. Recent efforts to overcome this problem via intermediate
steps have not been considered for this implementation \cite{Wegener2018}.

The quantum trajectories run with $N_{\mathrm{QT}} = 500$ and yield
an error above the MPDO and LTPN result, see again
Fig.~\ref{fig:otn:excitonconv}(b). While the error in MPDO and
LTPN is solely due to truncation and the error induced by the
Trotter decomposition, the QTs could further improve by taking more
trajectories. The error follows the law of large numbers. If we consider
the CPU times for each method obtained on a \mioseventeenx{} node compiled
with \emph{ifort}, the QTs exceed with $N_{\mathrm{QT}} = 500$ already
the resources for the MPDO. Here, QTs run with TEBD and, thus, the fastest
method. Therefore, more trajectories are not considered as they would
increase the difference in CPU time further. This CPU time is
the cumulative time for all simulations, and large-scale parallelization can make
them favorable for getting a quick picture of the physics. We point out once
again the very low bond dimension used in the MPDOs, which makes them
favorable, and models with higher entanglement might profit more from QTs.
This statement is supported by the next example, i.e.,
Sec.~\ref{sec:otn:conv:oqsb}. We recall that converting an MPS to an MPDO
increases the bond dimension from $\chi$ to $\chi^2$ and any decomposition,
such as an SVD, scales cubically with the matrix dimension.

\subsection{Local Lindblad operators with symmetry (QT and MPDO)               \label{sec:otn:conv:oqsb}}

The last convergence study treats a Lindblad master equation with a
conserved quantity. This symmetry restricts us to the usage of QTs and
MPDOs. We choose as our Hamiltonian the Bose-Hubbard model
\begin{eqnarray}                                                                \label{eq:otn:ham:bosehubbard}
  H &=& - J \sum_{k=1}^{L-1} \left( b_{k}^{\dagger} b_{k+1} + h.c. \right)
        + \frac{U}{2} \sum_{k=1}^{L} n_k (n_k  - 1)
        + \sum_{k=1}^{L} V(k) n_{k} \, .
\end{eqnarray}
$J$ sets the tunneling strength and the repulsive on-site
interaction is $U$. $b_{k}$ ($b_{k}^{\dagger}$) is the bosonic annihilation
(creation) operator acting on site $k$. The Hamiltonian conserves the number
of bosons, i.e., the commutator $ \left[ H, N \right] = 0$ with
$N = \sum_{k=1}^{L} n_k$. The chemical potential, i.e.,
$\mu \sum_{k=1}^{L} n_{k}$ can be discarded in simulations with number
conservation; it represents an energy shift without influence on the
simulation. But we include a potential $V(x)$ to capture the potential
of the double well. We choose a dephasing operator as the Lindblad operator,
i.e., $L_{k} = n_{k}$ acting on each site $k$.

\begin{figure}[t]
 \begin{center}
   \vspace{0.7cm}
   \begin{minipage}{0.47\linewidth}
      \vspace{0.05cm}
      \begin{overpic}[width=1.0 \columnwidth,unit=1mm]{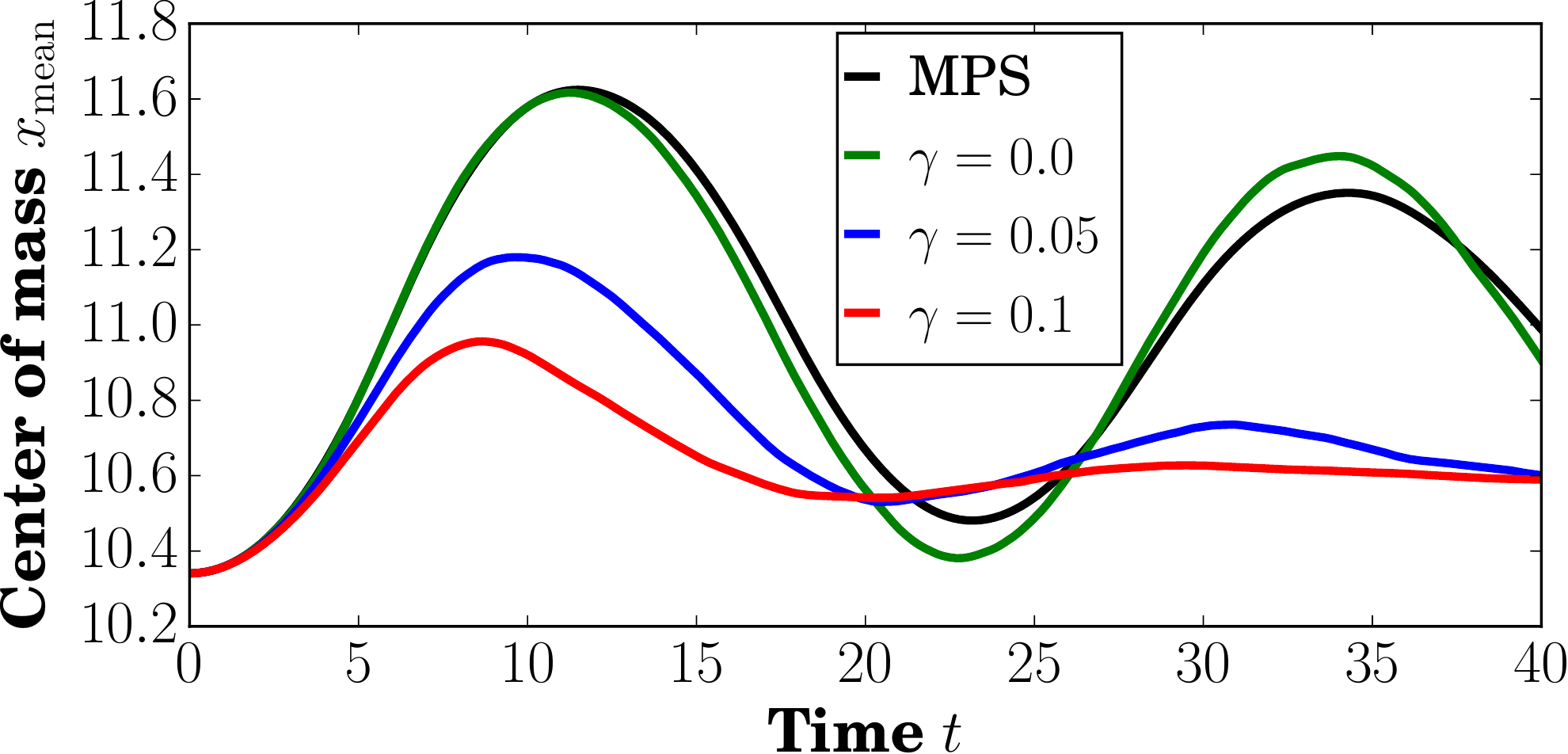}
        \put(2, 54){(a)}
      \end{overpic}
    \end{minipage}\hfill
    \begin{minipage}{0.47\linewidth}
      \vspace{0.05cm}
      \begin{overpic}[width=1.0 \columnwidth,unit=1mm]{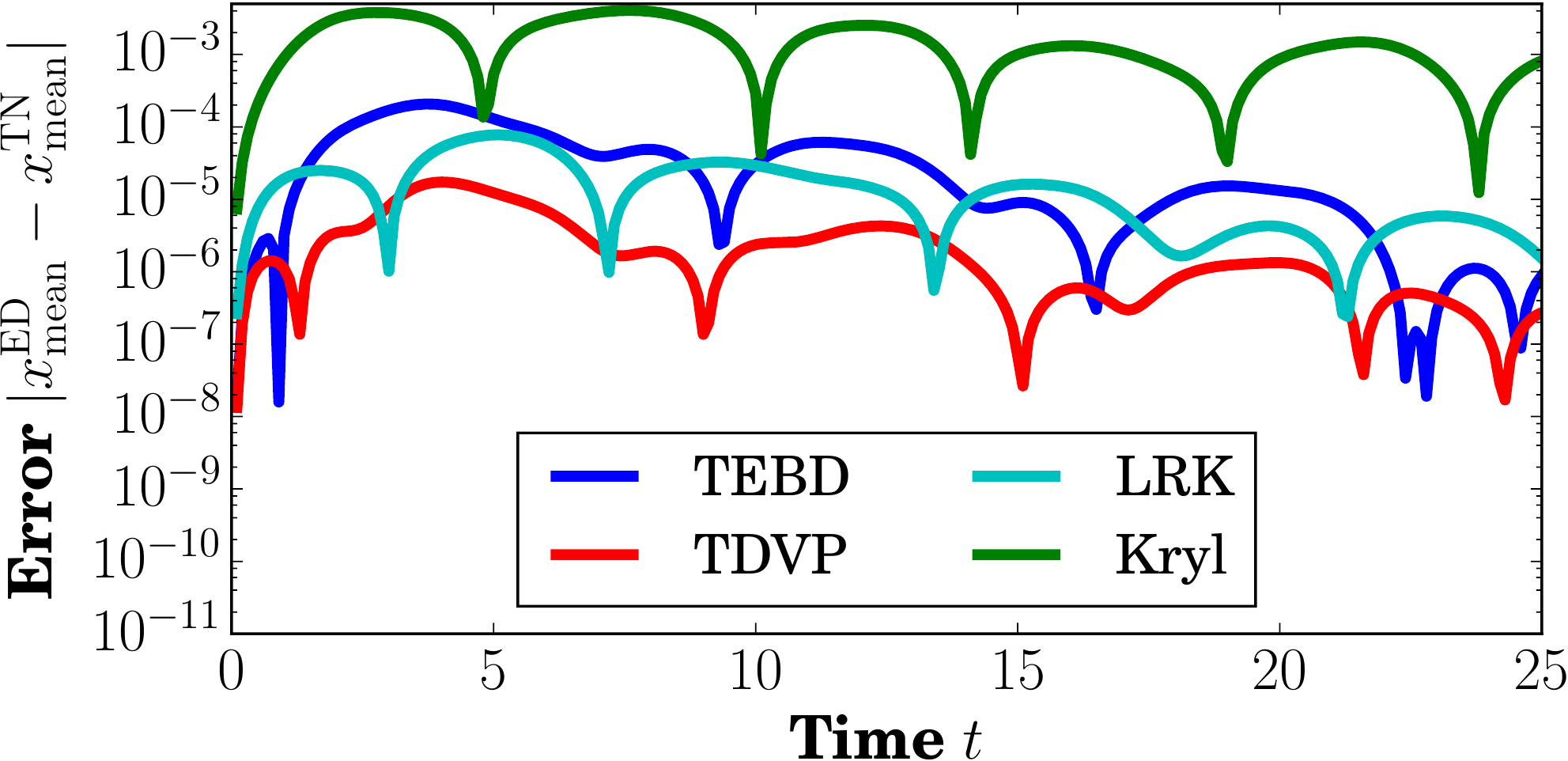}
        \put(2, 54){(b)}
      \end{overpic}
    \end{minipage}\vspace{0.5cm}
   \begin{minipage}{0.47\linewidth}
      \vspace{0.05cm}
      \begin{overpic}[width=1.0 \columnwidth,unit=1mm]{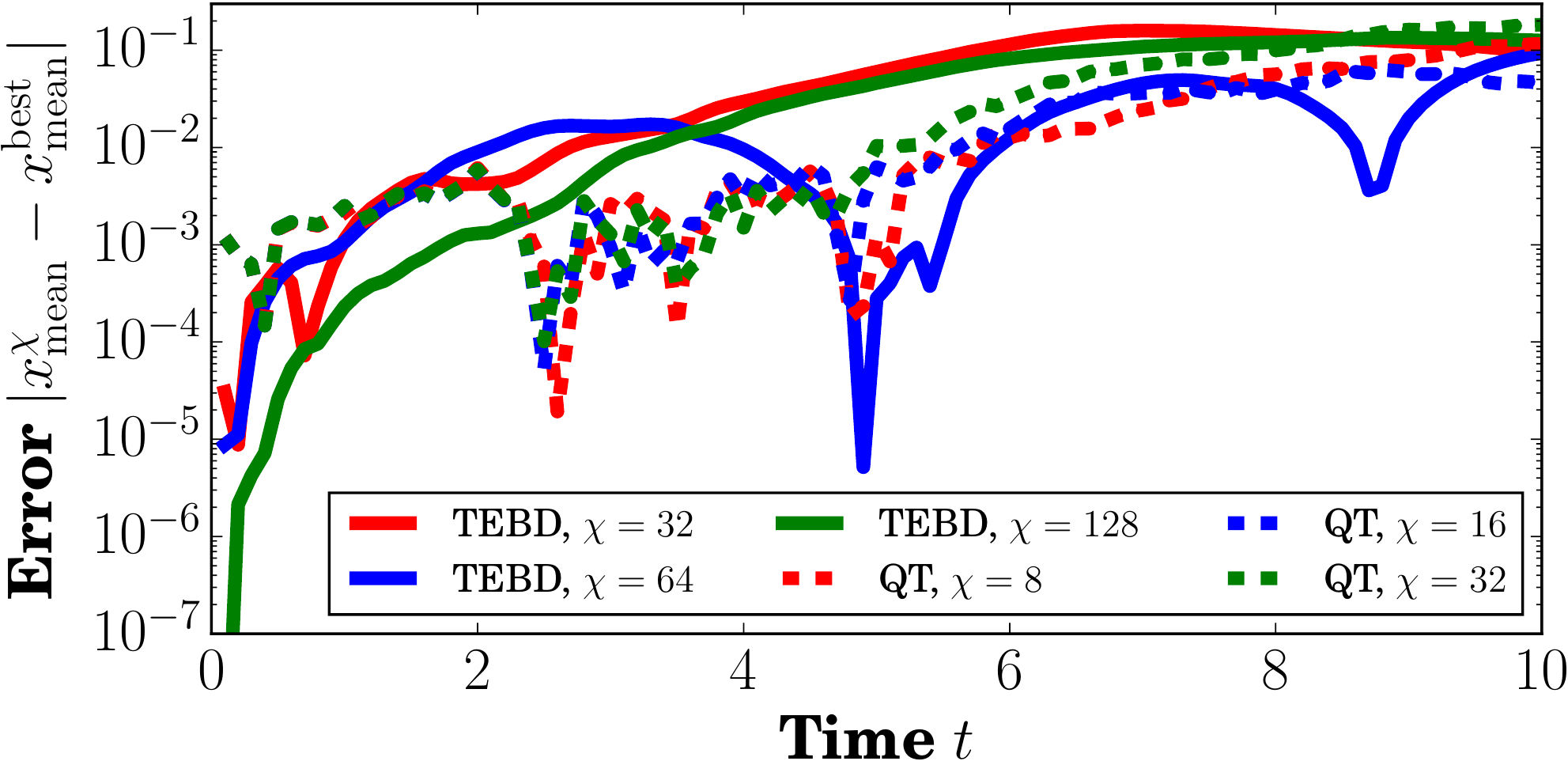}
        \put(2, 54){(c)}
      \end{overpic}
    \end{minipage}\hfill
    \begin{minipage}{0.47\linewidth}
      \vspace{0.05cm}
      \begin{overpic}[width=1.0 \columnwidth,unit=1mm]{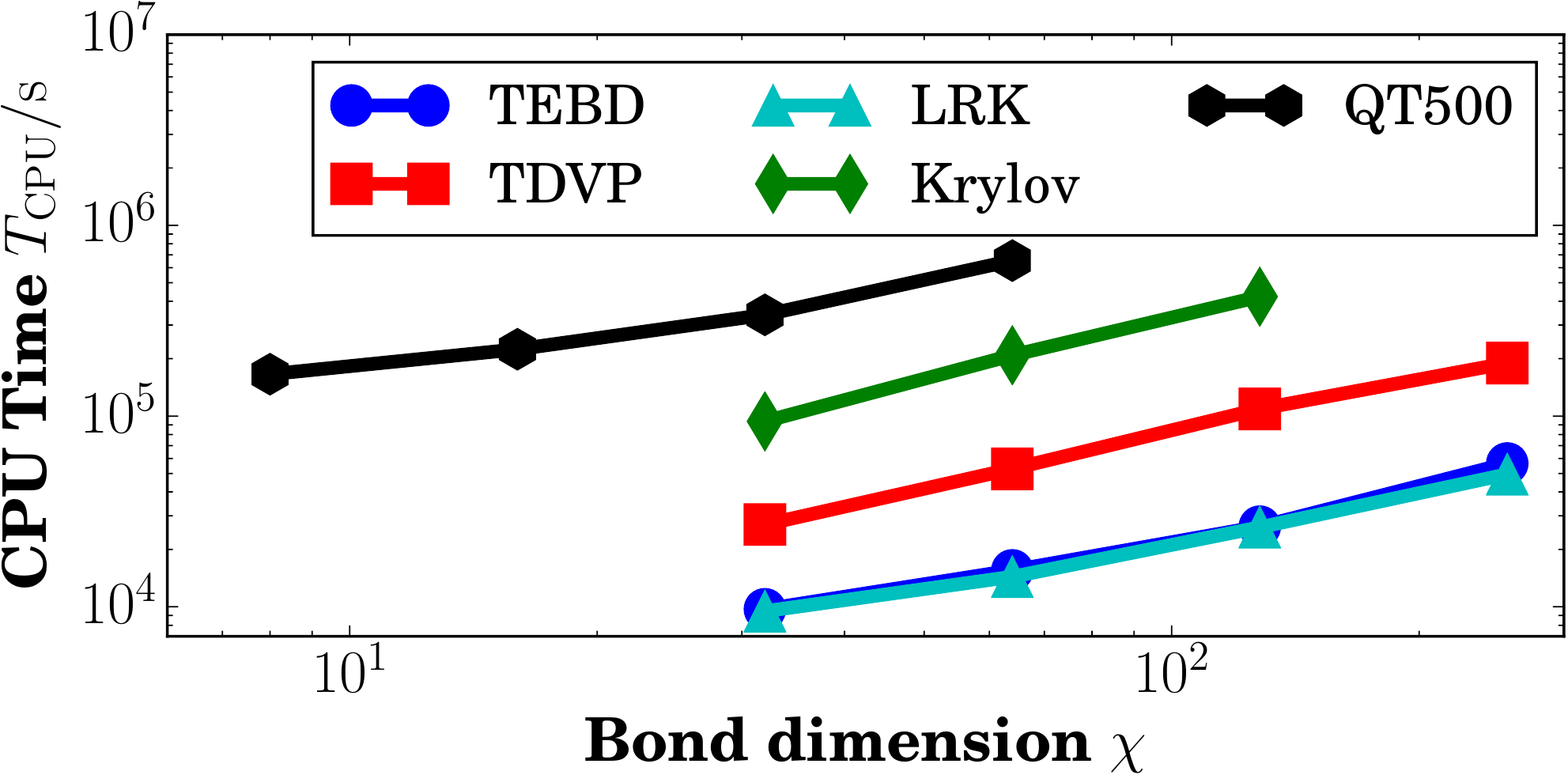}
        \put(2, 54){(d)}
      \end{overpic}
    \end{minipage}
   \caption[Convergence of open quantum system Bose-Hubbard dynamics]
   {\emph{Convergence of open quantum system Bose-Hubbard dynamics.}
     (a)~The center of mass starts oscillating after we release the
     potential on the right well at time $t = 0$ and keep only a
     one-site potential in the middle to separate wells. We observe
     that the simulations for MPS and MPDO of a closed system mismatch
     after about ten time units. The coupling of the system to a dephasing
     damps the oscillations.
     (b)~We compare the error of tensor network (TN) methods against exact
     diagonalization for a small system of $L = 5$.
     (c)~We compare the errors of QTs with TEBD and $N_{\mathrm{QT}} = 500$
     and simulations with MPDOs and TEBD to the TEBD simulations with the
     highest bond dimension, i.e., $\chi = 256$. Errors grow fast and
     show the numerical challenges associated with this problem setup.
     (d)~We compare the computational resources for MPDO and QT
     algorithms. The coupling to the environment is $\gamma = 0.05$.
     TEBD and LRK use the least resources, then comes TDVP. The quantum
     trajectories with $N_{\mathrm{QT}}$ and TEBD are located above MPDOs
     with TEBD. Due to the limitation in bond dimension in this example
     and the possibility of parallelizing, they are an attractive option.
                                                                                \label{fig:otn:doublewell}}
 \end{center}
\end{figure}

We consider a double-well scenario with two wells of $L_{W}$ sites, separated
by a one-site barrier with $V(L_{W} + 1) \neq 0$.
The total size of the system is $L = 2 L_{W} + 1$ and we
choose a filling of $N = L_{W}$. The initial state for the time evolution is
the ground state of the particles trapped in the left well with $V(x) = 1$
($V(x) = 0.5$) for $L = 5$ ($L = 21$) 
for all $x > L_{W}$. At $t = 0$, the wave function starts to oscillate
between the two wells in a Josephson-like regime. The coupling to the open
system damps out the oscillations; we show this effect in
Fig.~\ref{fig:otn:doublewell}(a) with the center of mass, or mean, as a function
of time, i.e.,
\begin{eqnarray}
  x_{\mathrm{mean}}(t) &=& \frac{1}{N} \sum_{k=1}^{L} k n_{k}(t) \, .
\end{eqnarray}
Two of the curves show the simulation for closed system time evolution with MPS and MPDO and
validate that the MPDO tensor network can simulate the closed system
correctly until entanglement generation introduces significant errors to
the MPDO. The effects of the dephasing lead to a damping of the
oscillation as shown in the two additional curves with $\gamma = 0.05$ and
$\gamma = 0.1$. The system size is $L = 21$, and we choose an initial bond
dimension of the ground state of $\chi = 5$. We assume that the ground
state has initial errors, but as we compare time evolution methods, all
time evolution methods start with the same ground state. As the initial
bond dimension of the MPDO is $\chi^2 = 25$, that allows us to run with
medium bond dimensions and judge on the error due to truncation. The data
shown in Fig.~\ref{fig:otn:doublewell}(a) is for bond dimension
$\chi = 256$ for the MPS and MPDO during the time evolution.

We presented the error for the exciton dynamics based on the exponential
decay; we are not aware that we have an equally suited result to compare
the dynamics of the double well and Bose-Hubbard model. Thus, we start
with a comparison between the tensor network algorithms and exact
diagonalization. The OSMPS package includes exact diagonalization methods
implemented in Python \cite{JaschkeED}, and one of their purposes is exactly
the accessibility for validating other techniques with an independent method.
We simulate with $L = 5$ sites and set the local dimension to $d = 3$,
i.e., states with more than two bosons per site are truncated. The error
of the center of mass for the simulation with TDVP is minimal through
most of the simulation and bounded by $10^{-5}$ according to
Fig.~\ref{fig:otn:doublewell}. The next best methods are LRK and TEBD,
while Krylov has the largest error bounded by $10^{-3}$. The choice of
the position of the center of mass a as macroscopic observable comes
with a couple of local minima of the error; the position of the center
of mass can align in two states, although the states are different and
the envelope is more meaningful and conclusive.

We now analyze how the convergence scales with bond dimension. The trends
are similar for the different time evolution methods, where the reason is
the exhausted bond dimension and truncation of singular values.
Figure~\ref{fig:otn:doublewell}(c) shows the error in the center of mass
of TEBD simulations with bond dimension $\chi = 32, 64, 128$ to the most
accurate simulation with $\chi = 256$. The significant difference between
the simulations reveals the need for high bond dimensions in this problem
and we cut the error plots at ten time units. We conclude that the bond
dimension is used completely after very short times for MPDOs; only
simulations with $\chi =128$ can keep the error below $10^{-5}$ for
the first couple of time steps.
The QTs averaged over $500$ runs are compared to the most accurate MPDO
simulation, both run with TEBD. We have three regimes. (i) The MPDO
simulations are more accurate at the very beginning of the time evolution
as the QTs suffer from statistical averaging. This regime ends at
$t \approx 2$ (ii) The error of the QTs drops below the MPDO simulations
with small bond dimension. The error due to truncation in the MPDO is
worse than the error from the average of the trajectories. (iii) For
times $t > 5$, the difference between the QTs and the MPDO with
$\chi = 256$ grows. Without involving more details, the difference can
be due to either method.

Finally, we turn to the actual resources used in
Fig.~\ref{fig:otn:doublewell}(d). We observe that both TEBD and LRK use
the same amount of resources. TDVP is less than an order of magnitude
above these two methods. The Krylov method takes much longer than TDVP,
which is in the end about an order of magnitude difference to the fastest
method, i.e., LRK. We run the QTs with a much
lower bond dimension; the actual bond dimensions $8$ ($16$) are comparable
after squaring it in the transformation from MPS to an MPDO of bond
dimension $64$ ($256$). The resources are more than a magnitude higher; we
recall that they can be easily parallelized in contrast to an MPDO
simulation. The large error at low bond dimension for MPDOs makes QTs much
more attractive than in the previous case of the exciton transport.

\section{Conclusions                                                           \label{sec:otn:conclusion}}

%
In this work, we have presented three different evolution techniques for
one-dimensional quantum systems under the Lindblad master equation with tensor
network methods. Therefore, the \emph{Open Source Matrix Product States}
(OSMPS) package allows the quantum community to explore
the advantages of each method. Open quantum system dynamics provide access
to more advanced simulations of quantum simulators including the effects
of the interaction with the environment as simulated with the Lindblad
master equation throughout this work.

%
We described the bond dimension of the Liouville operator represented as
a matrix product operator (MPO); this MPO is used to evolve matrix product
density operators beyond nearest-neighbor interactions and the time-evolving
block decimation (TEBD), i.e., the time-dependent variational principle,
the local Runge-Kutta time evolution, and the Krylov time evolution.
Section~\ref{sec:otn:mpdo} provides insights into the
scaling of methods due to the bond dimension of the MPO.
For example, non-local rules with
a bond dimension $\chi_{\mathrm{H-MPO}}$ to represent the Hamiltonian $H$ have a bond dimension of
$\chi_{\mathcal{L}-\mathrm{MPO}} = 2 \chi_{\mathrm{H-MPO}}$ to depict the Liouville operator. If we define
Lindblad terms analog to Hamiltonian terms, e.g., we replace a nearest-neighbor interaction of a Hamiltonian with a Lindblad operator acting on nearest-neighbor sites, they have a bond dimension three
times as big as $\chi_{\mathrm{H-MPO}}$. We found that different approaches are
favorable in different situations. Locally purified tensor networks (LPTNs)
had the best scaling in the example
of the finite temperature states of the quantum Ising model. For the
dissipative dynamics of an exciton, we found MPDOs as the optimal solution due
to the low entanglement throughout the simulation. When the entanglement
generated during the time evolution exhausts the maximal bond dimension as
observed in the Bose-Hubbard double well, quantum trajectories (QTs)
became a good alternative to MPDOs.

The examples also show the problem sizes one can treat with the OSMPS
package. The finite temperature states via imaginary time evolution and
TEBD can treat almost as many sites
as imaginary time evolution with MPS can, i.e., finding the ground state.
In detail, we compared the thermal states for $L = 100$ sites. Changing
the local dimension or the final temperature can increase the computational
scaling away from the MPS case; the number of time steps necessary
scales inversely with the final temperature at a fixed time step in the
evolution. CPU times on the order of $1000$ seconds on a cluster
did not reach the limits of the method.

The exciton dynamics represents an example which does not generate much
entanglement during the evolution, and the simulation stays below $100$
seconds on an HPC cluster, given the best choice of the method being
TEBD with MPDOs. This system is evolved for $30$ time units of the local
potential with $600$ time steps and $80$ sites. Bond dimension is not
exhausted in the example, where the maximum is set to $512$. Admittedly,
this system can be scaled up more for research, which was prevented in
our case due to the exploration of the methods not scaling favorably.

In contrast, the double well simulation shows the effect of entanglement
generated during the time evolution and preventing more precise simulations
of larger systems. We have $21$ sites with $10$ bosons and a local dimension
of $4$. The MPDO simulations produce visible errors beginning at time $10$
in the units of the tunneling at a bond dimension of $\chi = 256$. The
dissipative Bose-Hubbard model with two-site Lindblad operators using
TEBD-MPDO follows the same direction with limited system sizes and further
restriction with regards to quantum trajectories and the construction of
MPOs.

However, we strongly emphasize that each of the examples presented handles
system sizes which are well beyond the means of exact diagonalization and,
therefore, tensor networks are a very fruitful possibility to numerically
explore these systems in the open quantum dynamics context.

%
These studies underline the necessity to explore different methods for the
simulation of density matrices with tensor networks and choose the most
suitable method for pushing the limits in each case. We have presented
useful case studies with the quantum Ising model at finite temperature,
dissipative exciton dynamics, and the open quantum system Bose-Hubbard
modeling a double well with dephasing. Each of them could be a starting point
for more extensive studies or serve as a blueprint to study other models,
e.g., finite temperature diagrams of other spin models of the $XY\!Z$ class.

Furthermore, the adaption of tensor network methods for open quantum
systems beyond one-dimensional cases is another goal for the research
community. The adaption of projected entangled pair states, i.e., the
two-dimensional analog of a matrix product state, to open quantum systems
in \cite{Kshetrimayum2017} has established the foundation for such an
approach. These projected entangled pair operators can now be considered
for further research applications.

Evidently, tensor network methods have to evolve past the Markovian case of
the Lindblad master equation. The development of methods within the tensor
network algorithms to capture such non-Markovian effects is one future
direction in the design of new methods.

\section*{Acknowledgments}

We gratefully appreciate contributions from and discussions with D.~Alcala,
I.~de~Vega, M.~T.~Lusk, G.~Shchedrin, and M.~L.~Wall. This work has been
supported by the AFOSR under grant FA9550-14-1-0287, and the NSF under
the grants PHY-1520915 and OAC-1740130. We acknowledge support of the U.K.
Engineering and Physical Sciences Research Council (EPSRC) through the
``Quantum Science with Ultracold Molecules" Programme
(Grant No. EP/P01058X/1). The calculations were carried out using the high
performance computing resources provided by the Golden Energy Computing
Organization at the Colorado School of Mines. S.M. gratefully acknowledges
the support of the DFG via a Heisenberg fellowship and the TWITTER project.


%

\appendix

\section{Non-Local Lindblad operators with symmetry (MPDO)                     \label{sec:otn:conv:oqsc}}

We discuss the case of a non-local Lindblad operator acting on two
neighboring sites and explain the challenges such equations face. The
Lindblad operators conserve the number of bosons. We choose as a Hamiltonian
the Bose-Hubbard model, but in contrast to Eq.~\eqref{eq:otn:ham:bosehubbard}
we do not have an on-site potential,
\begin{eqnarray}                                                                \label{eq:otn:ham:bosehubbard}
  H &=& - J \sum_{k=1}^{L-1} \left( b_{k}^{\dagger} b_{k+1} + h.c. \right)
        + \frac{U}{2} \sum_{k=1}^{L} n_k (n_k  - 1) \, .
\end{eqnarray}
$J$ sets the tunneling strength and the repulsive on-site
interaction is $U$. $b_{k}$ ($b_{k}^{\dagger}$) is the bosonic annihilation
(creation) operator acting on site $k$. The Hamiltonian conserves the number
of bosons, i.e., the commutator $ \left[ H, N \right] = 0$ with
$N = \sum_{k=1}^{L} n_k$. The chemical potential, i.e.,
$\mu \sum_{k=1}^{L} n_{k}$ can be discarded in simulations with number
conservation; it represents an energy shift without influence on the
simulation.
The dissipative state preparation for the Bose-Hubbard model is the focus
of the references \cite{Diehl2008,Kraus2008} and further discussed in
\cite{Mueller2012,Kordas2015}. We use the Lindblad operators from these
approaches, i.e.,
\begin{eqnarray}                                                                \label{eq:otn:bh:origlindblads}
  L_{k} &=& \left( b_k^{\dagger} + b_{k+1}^{\dagger} \right)
            \left( b_{k} - b_{k+1} \right) \, ,
            \quad k \in \{ 1, \ldots, (L - 1) \} \, .
\end{eqnarray}
This formulation is not directly suitable for any of our Lindblad rule
sets. In fact, the two-site operator is not a product term and is therefore
challenging. We recall that a Lindblad operator of the type $L = A + B$
cannot be split into two Lindblad operators $L_{A} = A$ and $L_{B} = B$;
a look at the corresponding terms of $L \rho L^{\dagger}$ reveals the
missing cross-terms. Thus, we create a new type of rule set which is
similar to the many-body string Lindblad operators, but allows us to
define different operators for the operators on the left and right:
\begin{eqnarray}                                                                \label{eq:otn:bh:lindblads}
  L_{A,k \cdots k'} \rho L_{B,k \cdots k'}^{\dagger}
  - \frac{1}{2} \left\{ L_{B,k \cdots k'}^{\dagger} L_{A,k \cdots k'} ,
  \rho \right\} \, ,\\
  L_{A,k \cdots k'} = L_{A,k} \otimes L_{A,k+1} \otimes \cdots \otimes L_{A,k'} \, , \\
  L_{B,k \cdots k'} = L_{B,k} \otimes L_{B,k+1} \otimes \cdots \otimes L_{B,k'} \, .
\end{eqnarray}
Now, we can define the following set of Lindblad operators $L_{\mu}$
\begin{eqnarray}
  L_{\mu=1,k} &=& n_{k} \, ; \quad
  L_{\mu=2,A,k}  =  n_{k} \, , L_{\mu=2,B,k} = - b_{k}^{\dagger} b_{k+1} \, ; \nonumber \\
  L_{\mu=3,A,k} &=& n_{k} \, , L_{\mu=3,B,k} =   b_{k} b_{k+1}^{\dagger} \, ; \quad
  L_{\mu=4,A,k}  =  n_{k} \, , L_{\mu=4,B,k} = - n_{k+1} \, ; \nonumber \\
%
  L_{\mu=5,A,k} &=& - b_{k}^{\dagger} b_{k+1} \, , L_{\mu=5,B,k} = n_{k} \, ; \quad
  L_{\mu=6,A,k}  =  - b_{k}^{\dagger} b_{k+1} \, , L_{\mu=6,B,k} = - b_{k}^{\dagger} b_{k+1} \, ; \nonumber \\
  L_{\mu=7,A,k} &=& - b_{k}^{\dagger} b_{k+1} \, , L_{\mu=7,B,k} = b_{k} b_{k+1}^{\dagger} \, ; \quad
  L_{\mu=8,A,k}  =  - b_{k}^{\dagger} b_{k+1} \, , L_{\mu=8,B,k} = - n_{k+1} \, ; \nonumber \\
%
  L_{\mu= 9,A,k} &=& b_{k} b_{k+1}^{\dagger} \, , L_{\mu= 9,B,k} = n_{k} \, ; \quad
  L_{\mu=10,A,k}  =  b_{k} b_{k+1}^{\dagger} \, , L_{\mu=10,B,k} = - b_{k}^{\dagger} b_{k+1} \, ; \nonumber \\
  L_{\mu=11,A,k} &=& b_{k} b_{k+1}^{\dagger} \, , L_{\mu=11,B,k} = b_{k} b_{k+1}^{\dagger} \, ; \quad
  L_{\mu=12,A,k}  =  b_{k} b_{k+1}^{\dagger} \, , L_{\mu=12,B,k} = - n_{k+1} \, ; \nonumber \\
%
  L_{\mu=13,A,k} &=& n_{k+1} \, , L_{\mu=13,B,k} = n_{k} \, ; \quad
  L_{\mu=14,A,k}  =  n_{k+1} \, , L_{\mu=14,B,k} = - b_{k}^{\dagger} b_{k+1} \, ; \nonumber \\
  L_{\mu=15,A,k} &=& n_{k+1} \, , L_{\mu=15,B,k} = b_{k} b_{k+1}^{\dagger} \, ; \quad
  L_{\mu=16,k}  =  n_{k+1} \, .
\end{eqnarray}
The representation of this set of Lindblad operators together with the
Hamiltonian in an MPO is not recommended. We have a bond dimension of
$48$ (2 for local terms, 4 for tunneling in the Hamiltonian, $3 \times 14$
for the nearest-neighbor Lindblad terms). From the perspective of this bond
dimension, only TEBD remains as a valid option for time evolution with MPDOs.
QTs have problems due to the different terms $L_{A,k \cdots k'}$ and
$L_{B,k \cdots k'}$. Therefore, extension of the QTs to Lindblad operators
with no convenient representation in terms of rule sets, such as the example
in Eq.~\eqref{eq:otn:bh:origlindblads}, are convenient and considered for
the future.

The coupling of the single-site Lindblad operator $n_{k}$ is a function
of the lattice site; the two boundary sites have only half the coupling
of the bulk. We point
out that each of the terms conserves the number of bosons in the system and
thus the Liouville operator commutes with the global number operator
$N$, i.e., $\left[ \mathcal{L}, N \otimes \1 + 1 \otimes N^T \right] = 0$.
We defined this condition in Eq.~\eqref{eq:otn:liouvillesymm}.

For the simulation, we start in the ground state of the Bose-Hubbard model
in the Mott insulating phase at unit filling. The BKT transition is around
$J / U \approx 0.305$ \cite{Ejima2011} in the thermodynamic limit. We
choose $J = 0.1$ and
$U = 1.0$ is the energy scale. At $t=0$, we couple the system to the
reservoir with the Lindblad operators defined in Eq.~\eqref{eq:otn:bh:lindblads}.
We define the
depletion as $\xi = 1 - \max_{i} \left( \Xi_{i} \right) / \sum_{i} \Xi_{i}$
with $\Xi_{i}$ being the eigenvalues of the single-particle density matrix
$\langle b_{k}^{\dagger} b_{k'} \rangle$. The depletion decreases for the
superfluid as compared to the Mott insulator when dealing with ground states.

\begin{figure}[t]
 \begin{center}
   \vspace{0.7cm}
   \begin{minipage}{0.47\linewidth}
      \vspace{0.05cm}
      \begin{overpic}[width=1.0 \columnwidth,unit=1mm]{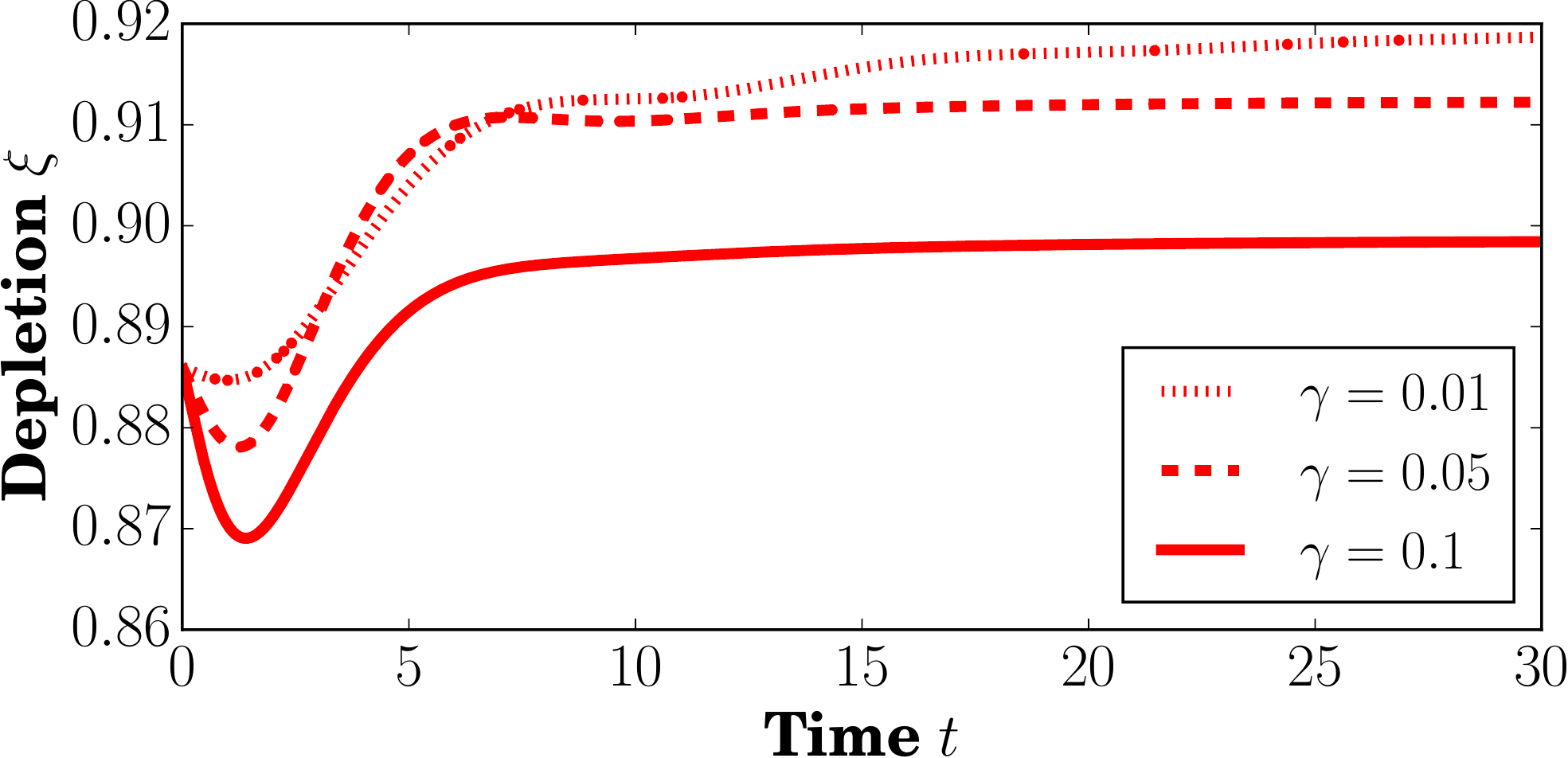}
        \put(2, 54){(a)}
      \end{overpic}
    \end{minipage}\hfill
    \begin{minipage}{0.47\linewidth}
      \vspace{0.05cm}
      \begin{overpic}[width=1.0 \columnwidth,unit=1mm]{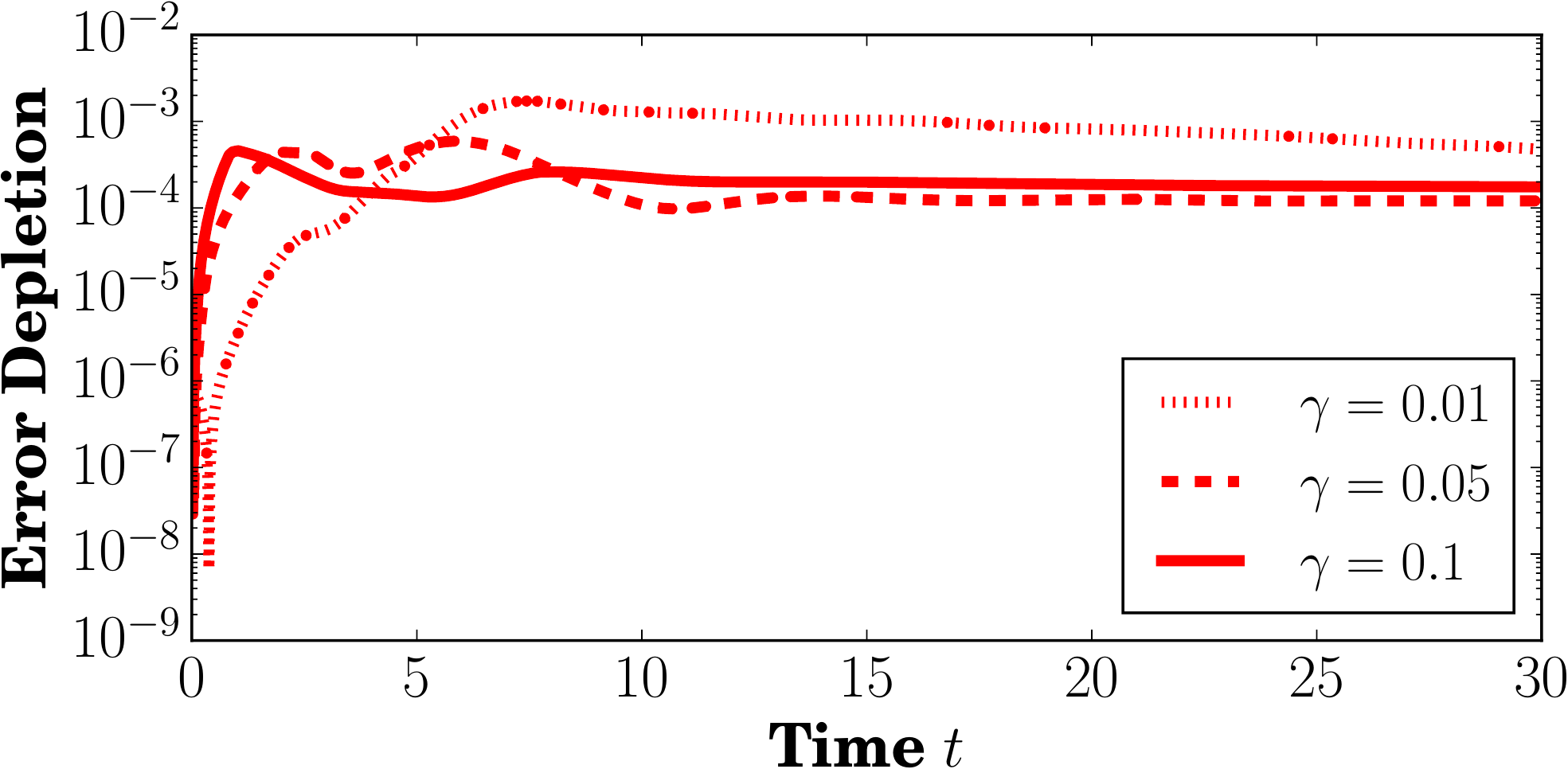}
        \put(2, 54){(b)}
      \end{overpic}
    \end{minipage}
   \caption[Non-local Lindblads with the Bose-Hubbard model]
   {\emph{Non-local Lindblads with the Bose-Hubbard model.}
     (a)~The dynamics of the depletion converging to the steady-state value
     with a two-site nearest-neighbor Lindblad operator.
     (b)~We calculate the dynamics for a small system $L = 5$ with exact
     diagonalization and TEBD and show the error of the TEBD as compared to exact
     diagonalization for the depletion with three different coupling strenghts
     $\gamma$.
                                                                                \label{fig:otn:superfluid}}
 \end{center}
\end{figure}

Figure~\ref{fig:otn:superfluid}(a) describes the evolution of the depletion
towards the steady state. We choose a system size of $L = 20$ and consider
states up to two bosons; we have a local dimension of $d = 3$. The data
is for a bond dimension of $\chi = 256$. The steady state differs for the
coupling strengths $\gamma = 0.01, 0.05, 0.1$ and has a lower depletion for
stronger coupling to the reservoir.
We now look at the convergence of the different methods; unfortunately, we
are not aware of a good observable to check as we had with the exponential
decay in the exciton example. Thus, we compare between the
exact diagonalization implementation and the tensor network algorithms in
Fig.~\ref{fig:otn:superfluid}(b). We consider a system size of $L = 5$ due
to the restriction of simulation a system in Liouville space with exact
diagonalization methods. We obtain that the value of the depletion is
converged at the order of $10^{-3}$ where the value of the depletion is on
the order of one. The bond dimension for the MPDO simulation is $\chi = 100$.

We conclude that the simulation of Lindblad operators acting on multiple sites
is possible. For such operators which cannot be represented as product term,
the simulation can get expensive to infeasible when MPOs and the corresponding
time evolution method are used. The TEBD algorithm, as shown in this
example, is the most convenenient alternative for such system dealing only with
nearest-neighbor interactions. We recall that TEBD is restricted to
nearest-neighbor terms in OSMPS.

\section{Bond dimension infinite-T Bose-Hubbard state                          \label{app:inftbh}}

We derive an upper bound for the bond dimension of the Bose-Hubbard model
in an MPDO for the
infinite temperature states. We assume that we have a local dimension $d$
and a filling of $N$ particles. We pick $L$ sufficiently large such that
$d \ll L$ and $N \approx L$. We pick a site in the middle of the system
and obtain $N + 1$ blocks in the block-diagonal structure. The first $d$
blocks have bond dimensions $1, 2, \ldots, d$ and the last $d$ blocks have
bond dimensions $d, d-1, \ldots, 2, 1$. The remaining fillings have a
bond dimension of $d$. Thus, we have a total number of subtensors
\begin{eqnarray}
  n &=& d (N - 2d + 1) + d (d + 1) = d (N - d + 2) \, ,
\end{eqnarray}
where the number of subtensors can serve as an upper bound to estimate
the bond dimension $\chi$, i.e., $\chi \le n$.

\end{document}